\documentclass[a4paper,onecolumn,11pt,unpublished]{quantumarticle} 
\pdfoutput=1

\usepackage[utf8]{inputenc}
\usepackage[english]{babel}
\usepackage[T1]{fontenc}
\AtBeginDocument{%
  {\fontencoding{T1}\fontfamily{lmr}\selectfont\fontfamily{lmss}\selectfont}%
  \DeclareFontShape{T1}{lmr}{m}{scit}{<->ssub*lmr/m/scsl}{}%
  \DeclareFontShape{T1}{lmss}{m}{sc}{<->ssub*lmr/m/sc}{}%
  \normalfont
}
\usepackage{amsmath}
\usepackage{hyperref}
\usepackage{amssymb}
\usepackage{graphicx}
\usepackage{braket}
\usepackage{algorithm}
\usepackage{algpseudocode}
\usepackage{float} 
\usepackage[dvipsnames]{xcolor}
\usepackage{amsthm}
\usepackage{qcircuit}
\usepackage{hyperref}
\usepackage{pgfplots}
\usepackage{tabularray}
\UseTblrLibrary{diagbox}

\usepackage[capitalise]{cleveref}
\usepackage{parskip}
\usepackage{comment}
\usepackage{multirow}
\usepackage{wrapfig}

\usepackage{tikz}
\usetikzlibrary{calc}
\usetikzlibrary{positioning}
\usetikzlibrary{decorations.pathreplacing,arrows.meta}

\usepgfplotslibrary{fillbetween}
\usepackage{pgfplots}
\pgfplotsset{compat=1.18}

\newtheorem{theorem}{Theorem}
\newtheorem{lemma}{Lemma}

\newtheorem{corollary}{Corollary}
\newtheorem{remark}{Remark}

\newcommand{\Paragraph}[1]{\smallskip\noindent{\bf #1}}

\newcommand{\ProofParagraph}[1]{\emph{#1}} 

\newcommand{\titlebox}[2]{
\begin{tikzpicture}
\node[draw, rectangle, thick,inner sep=4mm, align=left, text width=\textwidth-6mm] (titlebox) {#2};
\node[fill=white] (Title) at ($(titlebox.north) + (-0.18\textwidth, 0)$) {#1};
\end{tikzpicture}
}

\newcommand{\GateStyle}[1]{\textup{#1}}

\newcommand{\SGate}{\GateStyle{S}}
\newcommand{\HGate}{\GateStyle{H}}
\newcommand{\CXGate}{\GateStyle{CNOT}}
\newcommand{\XGate}{\GateStyle{X}}
\newcommand{\ZGate}{\GateStyle{Z}}
\newcommand{\YGate}{\GateStyle{Y}}
\newcommand{\IGate}{\GateStyle{I}}

\newcommand{\CZGate}{\GateStyle{CZ}}

\newcommand{\TGate}{\GateStyle{T}}

\newcommand{\TOF}{\GateStyle{TOF}}

\newcommand{\gateset}{\mathcal{G}}

\newcommand{\GenericSet}{\mathcal{S}}

\newcommand{\SId}{\GenericSet_{\IGate}} 
\newcommand{\SCliff}{\GenericSet_{\textup{Cliff}}} 

\newcommand{\GeneralPermutation}{\GenericSet_{\textup{GP}}} 

\newcommand{\ComplexityClass}[1]{\mathbf{#1}}
\newcommand{\PTIME}{\ComplexityClass{P}}
\newcommand{\NP}{\ComplexityClass{NP}}

\newcommand{\NQP}{\ComplexityClass{NQP}}
\newcommand{\QMA}{\ComplexityClass{QMA}}

\newcommand{\BQP}{\ComplexityClass{BQP}}
\newcommand{\PQP}{\ComplexityClass{PQP}}
\newcommand{\PSPACE}{\ComplexityClass{PSPACE}}

\newcommand{\PP}{\ComplexityClass{PP}}

\newcommand{\reals}{\mathbb{R}}

\newcommand{\FRACSAT}[1]{\gamma(#1)}

\newcommand{\Zring}[1]{\mathbb{Z}[#1]}
\newcommand{\Complex}{\mathbb{C}}

\newcommand{\Unitary}[1]{\mathcal{U}(#1)}

\newcommand{\FullDist}[2]{\mathcal{E}_{#2}(#1)}

\DeclareMathOperator{\Tr}{Tr}
\DeclareMathOperator{\diag}{diag}
\renewcommand{\Re}{\operatorname{Re}}
\renewcommand{\Im}{\operatorname{Im}}

\newcommand{\OperatorNorm}[1]{||#1||_\OperatorAbbr}

\newcommand{\Metric}[3]{d_{#3}(#1, #2)}

\newcommand{\BuresAngle}{\textup{BA}}

\newcommand{\SineMetric}{\textup{Sine}}
\newcommand{\OperatorAbbr}{\textup{op}}
\newcommand{\SpectralAbbr}{\textup{sp}}

\newcommand{\Problem}[1]{\textup{\textsc{#1}}}
\newcommand{\ATMOSTSAT}{\Problem{AtMostSat}}
\newcommand{\PROMISEATMOSTSAT}{\Problem{PromiseAtMostSat}}
\newcommand{\ApproxMember}{\Problem{ApproxMember}}
\newcommand{\PromiseApproxMember}{\Problem{PromiseApproxMember}}
\newcommand{\ApproxIdentity}{\Problem{ApproxId}}
\newcommand{\PromiseApproxIdentity}{\Problem{PromiseApproxId}}
\newcommand{\ApproxOpt}{\Problem{ApproxOpt}}

\definecolor{otublue}{RGB}{0,114,178}
\definecolor{otugreen}{RGB}{0,158,115}
\definecolor{otupurple}{RGB}{204,121,167}
\definecolor{otured}{RGB}{213,94,0}
\definecolor{otusky}{RGB}{86,180,233}

\title{On the Hardness of Approximate Quantum Circuit Membership and Optimization}

\author{Adam Husted Kjelstrøm}
\email{husted@cs.au.dk}
\affiliation{Department of Computer Science, Aarhus University}
\orcid{0000-0002-1275-9933}

\author{Jaco van de Pol}
\email{jaco@cs.au.dk}
\affiliation{Department of Computer Science, Aarhus University}
\orcid{0000-0003-4305-0625}

\author{Andreas Pavlogiannis}
\email{pavlogiannis@cs.au.dk}
\affiliation{Department of Computer Science, Aarhus University}
\orcid{0000-0002-8943-0722}

\begin{document}

\begin{abstract}
Quantum circuit optimization is a standard, indispensable component of the quantum software stack, used to reduce noise, maximize performance, alleviate costs of quantum error correction, or tailor quantum circuits to specific hardware platforms.
Another natural problem is the approximate membership question:~for a fixed set of unitaries $\GenericSet$, decide whether a given circuit $C$ can be approximated within a distance $\varepsilon$ by a unitary in $\GenericSet$, according to some metric $m$.
Membership is a special case of optimization, where $\GenericSet$ specifies a set of
unitaries that can be implemented at no cost (e.g., with noise-minimal circuits, or without the need for magic states).
In a well-known result, Janzing, Wocjan, and Beth (Int. J. Quantum Inf., 2005) showed that, for $m$ being the operator metric, the special case of membership where $\GenericSet=\{\IGate\}$ (i.e., approximate identity) is $\QMA$-complete when $C$ comes with a promise that it is either close to the identity or sufficiently far from it.

In this work, we generalize the setting of Janzing, Wocjan, and Beth in three natural directions:
(i)~we work over promise-free inputs, 
(ii)~besides the case $\GenericSet=\{\IGate\}$, we also consider the cases where $\GenericSet$ contains only Clifford unitaries, or only generalized permutation matrices (i.e., no superposition), and
(iii)~we measure unitary distances under several metrics $m$, namely, the trace-based, spectral, and operator metrics.
We show that, in all cases, the approximate membership question is hard even for $\PP$, for wide ranges of $\varepsilon$, provided $C$'s gate set can implement Clifford+$\TGate$, which is a property of the gate-sets provided by standard hardware.
In particular, if $\GenericSet=\{\IGate\}$, or if $\GenericSet$ contains only generalized permutation matrices, $\IGate\in\GenericSet$, and $m$ is a trace-based distance, then $\PP$-hardness holds for all possible (non-trivial) values of $\varepsilon$.
Moreover, if $\GenericSet=\{\IGate\}$ and $m$ is a trace-based distance, then $\PP$-hardness holds even for circuits $C$ over classical-reversible gate-sets.

Overall, our results indicate that the approximate membership problem is harder over promise-free inputs, compared to $\QMA$-completeness of Janzing, Wocjan, and Beth, and this hardness persists across quantum resources (any gates, non-Clifford, superposition) and various distance metrics.
Finally, these results directly imply the $\PP$-hardness of circuit optimization when minimizing total, non-Clifford, or superposition gate depth or count.
\end{abstract}

\maketitle

\section{Introduction}

Optimizing quantum circuits, that is, given a circuit $C$, obtaining an (approximately) equivalent circuit $C'$ that minimizes an objective function, is a standard task to improve the performance of quantum computers, thus producing circuits that are tailored to a given quantum run-time environment.
By picking different objective functions, optimization can target both NISQ and error-corrected platforms by minimizing, e.g., the number of multi-qubit gates or non-Clifford resources.
Permitting a bit of slack, i.e., allowing an output circuit that is a close (but not necessarily exact) match, is typical~\cite{Dawson2005solovay_kitaev_thrm,Kliuchnikov_2014,Kliuchnikov2016Practical,khatri2019quantum,Mukhopadhyay_2021,G_netay_Johansen_2021,Gheorghiu2022TCountTDepth,selinger2014efficientcliffordtapproximationsinglequbit,ross2016optimalancillafreecliffordtapproximation}.
This is partly because it gives the optimizer more freedom, which may yield better results, and partly because it is often unavoidable: for finite gate-sets such as Clifford+$\TGate$, universality holds only approximately, so most unitaries cannot be implemented exactly.
Circuit optimization is often phrased as a decision problem given a circuit $C$ and budget $k$. It is parameterized by a cost function $f$, a metric $m$, a tolerance $\varepsilon\geq 0$, and a gate-set $\gateset$, where $d_m(C, C')$ denotes the distance between the unitaries of $C$ and $C'$ under $m$:

\titlebox{$\ApproxOpt(f,m,\varepsilon, \gateset)$}{
\emph{Input:} A circuit $C$ over $\gateset$ and an integer $k \geq 0$. \\
\emph{Question:} Is there a circuit $C'$ over $\gateset$ s.t.\ $\Metric{C}{C'}{m} \leq \varepsilon$ and $f(C') \leq k$?
}

Another common computational problem is the (approximate) equivalence question:~given two circuits $C$ and $C'$, are their corresponding unitaries $\varepsilon$-close to each other, according to some metric $m$?
The equivalence question often serves as a subroutine of peephole optimization, whereby a
subcircuit $C$ can be replaced by a
circuit $C'$ with a small cost $f(C')$, provided that it encodes an equivalent computation.
The case where $C'$ is the empty circuit (asking whether $C$ is $\varepsilon$-close to the identity) has received special attention.
Indeed, besides being theoretically fundamental, this question also underpins practical circuit-reduction techniques, whereby parts of a circuit are deleted if they are deemed to be close enough to the identity~\cite{Barenco_1996,Escofet_2026,Yale_2025}.
When $C$ comes with a promise that its distance from $\IGate$ in the operator metric is either $\leq \varepsilon$ or $\geq \delta$ with $\delta-\varepsilon > n^{-O(1)}$, the resulting problem, known as non-identity check, is well known to be $\QMA$-complete over gate-sets $\gateset$ of arbitrary one- and two-qubit gates~\cite{Janzing2005nonIdentityCheckIsQMA_Complete}.
On the other hand, insisting on $\varepsilon=0$ 
corresponds to exact non-identity check, which is $\NQP$-complete for circuits $C$ over Clifford+$\TGate$~\cite{Tanaka2010ExactNonIdentityCheckNQPComplete,kjelstrom2025exactquantumcircuitoptimization,Wetering2023OptimisingQuantumCircuitsHard}.

A natural generalization of equivalence is the membership question. It is parameterized by a set of unitaries $\GenericSet$ that are considered to be of minimal cost. Instead of comparing with a single candidate circuit $C'$, the task is to decide whether $C$ is approximately equivalent to any element of $\GenericSet$.

\titlebox{$\ApproxMember(\GenericSet, m,\varepsilon,\gateset)$}{
\emph{Input:} A circuit $C$ over $\gateset$. \\
\emph{Question:} Is there a $U\in \GenericSet$ s.t.\ $\Metric{C}{U}{m} \leq \varepsilon$?
}

Approximate equivalence with the empty circuit is the special case $\GenericSet=\{\IGate\}$ of $\ApproxMember$, which we call $\ApproxIdentity$.
In this work, we consider two additional cases for $\GenericSet$, namely, $\GenericSet\subseteq \SCliff$ and $\GenericSet\subseteq \GeneralPermutation$, where $\SCliff$ is the set of unitaries implemented by Clifford circuits and $\GeneralPermutation$ is the set of generalized permutation matrices (which do not create superposition).
We further require that $\IGate\in \GenericSet$.
Each case asks whether $C$ can be approximated to precision $\varepsilon$ with restricted quantum resources: no gates at all ($\GenericSet=\{\IGate\}$), only Clifford gates ($\GenericSet\subseteq\SCliff$), or no superposition gates ($\GenericSet\subseteq\GeneralPermutation$).

We consider three standard metrics $m$ for defining the distance between two unitaries $U$ and $V$.
The Sine metric is defined as a function of the trace $|\Tr(U^\dagger V)|$, which captures the ``average-case'' difference between $U\ket{\psi}$ and $V\ket{\psi}$.
Related measures include the Bures angle, Bures metric~\cite{Zhihao2009fidelity}, and average gate distance measures~\cite{Nielsen2002}. 
These have seen extensive use in quantum circuit synthesis~\cite{Kliuchnikov_2014,Kliuchnikov2016Practical,khatri2019quantum,Mukhopadhyay_2021,G_netay_Johansen_2021,Gheorghiu2022TCountTDepth}.
The spectral metric, induced by the operator norm, captures the ``worst-case'' difference between $U\ket{\psi}$ and $V \ket{\psi}$.
This is the metric in which the Solovay-Kitaev theorem is stated~\cite{Dawson2005solovay_kitaev_thrm}; it is also used for quantum circuit optimization~\cite{selinger2014efficientcliffordtapproximationsinglequbit,ross2016optimalancillafreecliffordtapproximation}.
The operator metric, also considered in~\cite{Wetering2023OptimisingQuantumCircuitsHard}, is given by the minimum of the spectral metric between $U$ and $e^{i\alpha}V$ over $\alpha\in \reals$.
This is a natural choice as the global phase is physically irrelevant, potentially permitting better optimizations.

\Paragraph{Our contributions.}
Our main result is that $\ApproxMember(\GenericSet,m,\varepsilon,\gateset)$ is $\PP$-hard for every set of unitaries $\GenericSet$, metric $m$, and tolerance $\varepsilon$ covered by~\cref{table:all_problems_hardness2}, where $\varepsilon$ is $\PTIME$-time computable, i.e., computable to $b$ bits of precision in time $b^{O(1)}$.

\begin{table}[t]
\centering
\begin{tblr}{
  colspec = {|c||c|c|c|},
  row{1}  = {abovesep=4pt, belowsep=4pt},
}
\hline
\diagbox{$m$}{$\GenericSet$} & Identity & Any Clifford set & 
\begin{tabular}{l}Any generalized\\permutation set\end{tabular}\\
\hline
\hline
Trace ($\SineMetric$) & $\mathbf{(0,1)}$ & $(0,\sqrt{2}/4)$ & $\mathbf{(0,1)}$ \\
\hline
Spectral & $\mathbf{(0,2)}$ & $(0,\sqrt{2}/4)$ & $(0,\sqrt{2})$ \\
\hline
Operator & $\mathbf{(0,2)}$ & $(0,\sqrt{2}/4)$ & $(0,\sqrt{2})$ \\
\hline
\end{tblr}
\caption{
The open interval of $\PTIME$-time computable $\varepsilon$ for which \cref{theorem:approx_member_hardness} establishes $\PP$-hardness for $\ApproxMember(\GenericSet, m,\varepsilon,\gateset)$, per metric $m$ and set of unitaries $\GenericSet$, where the columns cover $\GenericSet=\{\IGate\}$, every $\GenericSet$ with $\IGate\in\GenericSet\subseteq\SCliff$, and every $\GenericSet$ with $\IGate\in\GenericSet\subseteq\GeneralPermutation$, respectively.
Boldface entries mark ranges equal to $(0,D_m)$, i.e., every $\varepsilon$ strictly between $0$ and the diameter $D_m$ of the metric.
}
\label{table:all_problems_hardness2}
\end{table}

\smallskip
\begin{theorem}
    \label{theorem:approx_member_hardness}
    Consider any gate-set $\gateset$ that exactly implements Clifford+$\TGate$.
    For each combination of a set of unitaries $\GenericSet$, metric $m$, and $\PTIME$-time computable $\varepsilon$ in the range given by the corresponding cell in~\cref{table:all_problems_hardness2}, we have that $\ApproxMember(\GenericSet, m, \varepsilon, \gateset)$ is $\PP$-hard.
\end{theorem}

In the context of optimization, given some cost function $f$, let $\GenericSet_f$ denote the set of unitaries $U_{C'}$ of circuits $C'$ with $f(C')=0$.
\cref{theorem:approx_member_hardness} yields the following corollary.

\smallskip
\begin{corollary}
    \label{corollary:approx_opt_hardness}
    Consider any gate-set $\gateset$ that exactly implements Clifford+$\TGate$.
    For each combination of a set of unitaries $\GenericSet$, metric $m$, and $\PTIME$-time computable $\varepsilon$ in the range given by the corresponding cell in~\cref{table:all_problems_hardness2}, and any $f$ for which $\GenericSet_f = \GenericSet$, we have  $\ApproxOpt(f, m, \varepsilon, \gateset)$ is $\PP$-hard.
\end{corollary}

A few consequences are worth highlighting, with $\varepsilon$ $\PTIME$-time computable throughout.
First, optimizing total gate depth or count is $\PP$-hard under each of the three metrics for every $\varepsilon$ strictly between $0$ and the diameter of the metric.
Second, the same holds for optimizing superposition gate depth or count, this time only for the Sine metric.
Third, optimizing non-Clifford (e.g., $\TGate$) gate depth or count is $\PP$-hard for $0 < \varepsilon < \sqrt{2}/4\approx 0.3536$.
This strengthens a result of Van de Wetering and Amy~\cite{Wetering2023OptimisingQuantumCircuitsHard} who, under the operator metric, established $\NP$-hardness for $0<\varepsilon<\sin(\pi/16)\approx 0.1951$, and also generalizes it to the other metrics.
Fourth, $\ApproxIdentity(\SineMetric, \varepsilon, \gateset)$, and $\ApproxOpt(f,\SineMetric,\varepsilon, \gateset)$ for $f$ the total gate depth or count, remain $\PP$-hard for all $0<\varepsilon<1$ even when $\gateset$ consists only of classical-reversible gates (e.g., $\XGate$, $\CXGate$, and $\TOF$), which cannot create superposition; see~\cref{rem:sine_hardness_classical_reversible}.

Our $\PP$-hardness results also generalize in two ways.
First, for any distance measure $m_2$ that is derivable from one of the considered metrics $m_1$, i.e., $\Metric{\cdot}{\cdot}{m_2} = g(\Metric{\cdot}{\cdot}{m_1})$ for some strictly increasing function $g$, the corresponding membership and optimization problems under $m_2$ are also $\PP$-hard, for the respective approximation range obtained under $g$ (see~\cref{lemma:equivalence_of_metrics}).
In particular, every Sine entry of~\cref{table:all_problems_hardness2} transfers to the Bures angle, the Bures metric, and the average gate distance at the correspondingly transformed threshold.
This covers, for instance, the distance measures used in~\cite{khatri2019quantum,Gheorghiu2022TCountTDepth}, and shows that minimizing total depth or count to within any non-trivial (measure-specific) $\varepsilon$ for any distance measure derived from the Sine metric is $\PP$-hard.
Second, the optimization problems remain $\PP$-hard even to approximate the minimum cost within any factor (constant or not), because the hardness holds already when the minimum cost is $0$ (i.e., the optimal circuit $C'$ has $f(C')=0$ and thus $U_{C'}\in \GenericSet_{f}$); see~\cref{rem:factor_approximation}.

\begin{figure}[b]
\centering
\begin{tikzpicture}[
  node distance=6mm and 2mm,
  box/.style={draw, rounded corners=2pt, align=center,
              inner sep=6pt, minimum width=38mm}
]
  \node[box] (atmostsat) {$\ATMOSTSAT$};
  \node[box, below=of atmostsat]  (atmost) {$\PROMISEATMOSTSAT$};
  \node[box, below=of atmost]  (pApprodID)   {$\PromiseApproxIdentity$};
  \node[box, below=of pApprodID]    (approxCliff)    {\cref{theorem:approx_member_hardness}, $\GenericSet\subseteq\SCliff$};
  \node[box, left=of approxCliff]    (approxID)    {\cref{theorem:approx_member_hardness}, $\GenericSet=\SId$};
  \node[box, right=of approxCliff]    (approxGenPerm)    {\cref{theorem:approx_member_hardness}, $\GenericSet\subseteq\GeneralPermutation$};

  \draw[->] (atmostsat) -- node[right] {\cref{lemma:hardness_promise_atmostsat}} (atmost);
  \draw[->] (atmost) -- node[right] {\cref{theorem:approx_identity_promise_hardness}} (pApprodID);

  \draw[->] (pApprodID.west) -| node[right, pos=0.84] {($\star$)} (approxID.north);
  \draw[->] (pApprodID) -- node[right] {\cref{lemma:reduction_promise_approx_id_to_approx_member}} (approxCliff);
  \draw[->] (atmost.east) -| node[right, pos=0.929] {\cref{lemma:template}} (approxGenPerm.north);

\end{tikzpicture}
\caption{The chain of reductions. 
$\ATMOSTSAT(\beta)$ is by folklore $\PP$-hard for every fixed efficiently computable $0<\beta<1$, and each arrow is a Karp reduction, where an arrow into the bottom row reduces to $\ApproxMember(\GenericSet, m, \varepsilon, \gateset)$ for the indicated $\GenericSet$, with $\IGate\in\GenericSet$ throughout.
Hardness for each $\GenericSet$ is obtained in one of three ways: via a reduction from a promise problem to its promise-free variant, as is the case for ($\star$); via the triangle inequality argument of~\cref{lemma:reduction_promise_approx_id_to_approx_member}, which is the route taken for $\GenericSet\subseteq\SCliff$; or via~\cref{lemma:template}, as is the case for $\GenericSet\subseteq\GeneralPermutation$. }

\label{fig:reduction-chain}
\end{figure}
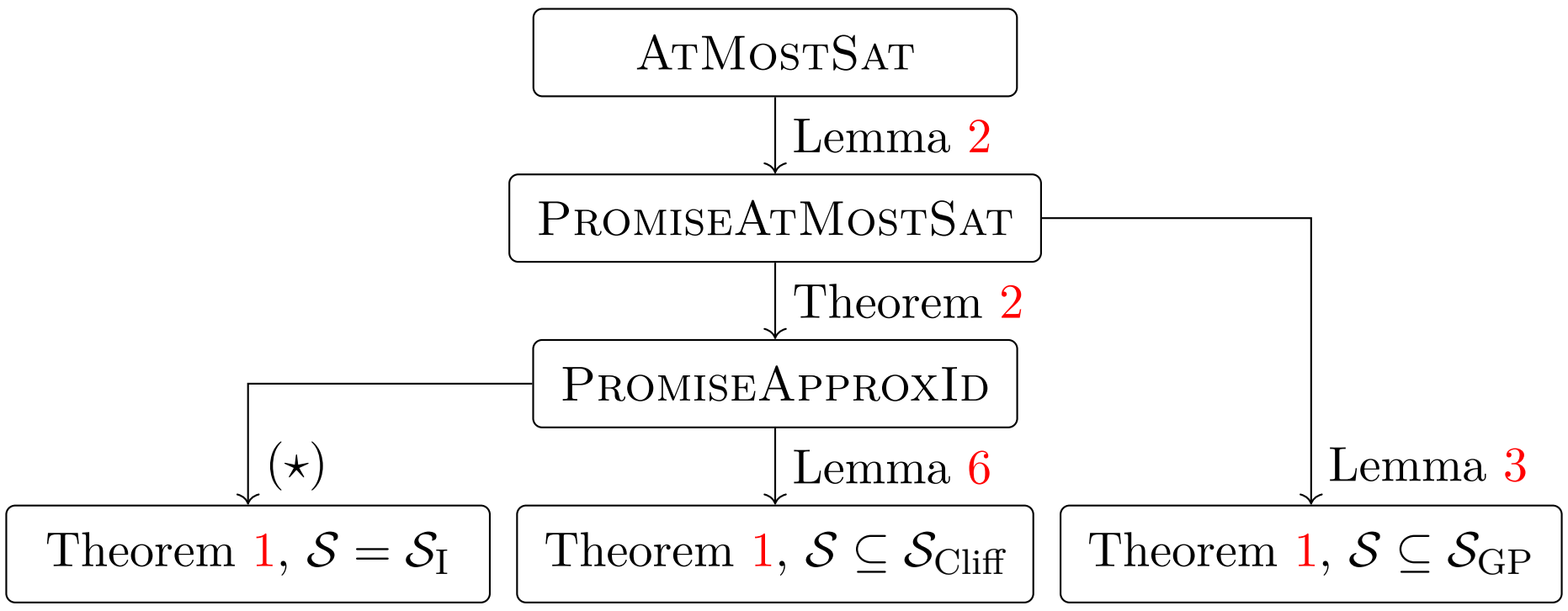

\Paragraph{Paper organization and proof structure.}
The remainder of the paper is organized as follows.
\cref{sec:definitions} fixes notation and recalls standard background on unitaries, quantum circuits, distance measures, and the class $\PP$.
\cref{fig:reduction-chain} shows the chain of reductions towards \cref{theorem:approx_member_hardness}.
We use $\ATMOSTSAT$ as the source of $\PP$-hardness, which asks whether a propositional formula $\varphi$ is satisfied by at most a fraction $\beta$ of its assignments.
In \cref{sec:framework_for_pp_hardness}, \cref{lemma:hardness_promise_atmostsat} shows that $\ATMOSTSAT$ remains $\PP$-hard under the promise that at most a fraction $\kappa$ of all assignments satisfy $\varphi$; we call this promise problem $\PROMISEATMOSTSAT$.
We then use the $\PP$-hardness of $\PROMISEATMOSTSAT$ to establish the hardness of $\PromiseApproxIdentity$ under all three metrics we consider (\cref{theorem:approx_identity_promise_hardness}).
In words, $\PromiseApproxIdentity$ is a promise version of $\ApproxIdentity$, asking whether a circuit $C$ is $\varepsilon$-close to $\IGate$ under the promise that it is $\tau$-close to it.
This naturally yields hardness for $\ApproxIdentity$ (as the two problems agree on every circuit satisfying the promise), and thus completes the case $\GenericSet=\{\IGate\}$ of \cref{theorem:approx_member_hardness}.
In \cref{sec:approx_member_instances_hardness}, we establish \cref{theorem:approx_member_hardness} for the cases $\IGate\in\GenericSet\subseteq\SCliff$ and $\IGate\in\GenericSet\subseteq\GeneralPermutation$.
We show that the hard circuits $C$ for $\PromiseApproxIdentity$ are also hard for $\ApproxMember(\GenericSet)$ with $\IGate\in\GenericSet\subseteq\SCliff$: intuitively, the promise that $C$ is always $\tau$-close to $\IGate$, combined with the triangle inequality of the respective metric, implies that the only elements of $\GenericSet$ that can be $\varepsilon$-close to $C$ are those at distance $0$ from $\IGate$, and thus $\ApproxMember(\GenericSet)$ coincides with $\PromiseApproxIdentity$ on circuits satisfying its promise, which is $\PP$-hard by \cref{theorem:approx_identity_promise_hardness}.
The case of $\IGate \in \GenericSet\subseteq \GeneralPermutation$ proceeds by reduction from $\PROMISEATMOSTSAT$: given $\varphi$, we construct circuits $C$ which are $\varepsilon$-close to $\IGate$ when $\varphi$ is satisfied by at most a fraction $\delta$ of its assignments, and at a distance greater than $\varepsilon$ from 
$\GenericSet$ otherwise (\cref{theorem:hardness_approx_non_superpos_sp_op,theorem:approx_non_superpos_hardness}).
We conclude in \cref{sec:conclusion} with open problems.

\section{Preliminaries} \label{sec:definitions}

In this section, we recall a number of standard definitions on unitary transformations, quantum circuits, metrics, and complexity theory.

\subsection{Unitary Transformations and Quantum Circuits}

\Paragraph{Unitary Transformations.}
Throughout, we write $U,V\in \Unitary{2^n}$ for $n$-qubit unitaries. A unitary (transformation) $U\in \Complex^{2^n\times2^n}$ satisfies $U U^\dagger=U^\dagger U = \IGate$.
By the spectral theorem, $U$ admits a factorization as $U=Q D Q^{\dagger}$, where $Q,D\in \Unitary{2^n}$ and $D$ is diagonal with the eigenvalues of $U$ on the diagonal.
The eigenvalues of $U$ are of the form $e^{i\phi}$ for real $\phi$; we call the set $\{\phi_j\}_j$ the eigenangles of $U$.

We write $U \equiv V$ if $U = e^{i\alpha} V$ for some $\alpha\in \reals$ and denote the tensor product of $U$ and $V$ by $U \otimes V$.
If $\{\phi_j\}_j$ and $\{\phi'_\ell\}_\ell$ are the eigenangles of $U$ and $V$, respectively, then the eigenangles of $U \otimes V$ are $\{\phi_j + \phi'_\ell\}_{j,\ell}$. 

\Paragraph{Gates and Quantum Circuits.} \label{subsec:quantum_circuits_to_unitaries}
A circuit $C$ is a finite sequence of quantum gates from some gate-set $\gateset$, each applied to specific (possibly different) qubits in a system. Each gate corresponds to a fixed unitary. We frequently use some standard gates, defined by their action on the basis qubits written $\ket{0}$ and $\ket{1}$; we let $a,b\in\{0,1\}$. The identity gate is $\IGate\ket{a}=\ket{a}$. The Pauli gates are defined as $\XGate\ket{a}=\ket{\neg a}$, $\ZGate\ket{a}=(-1)^a\ket{a}$ and $\YGate=i\XGate\ZGate$.

The Clifford gate-set consists of $\CXGate$, $\HGate$ and $\SGate$. Here the binary gate $\CXGate$ (Conditional NOT, entangling) is defined by $\CXGate\ket{a}\ket{b}=\ket{a}\ket{a\oplus b}$. The phase gate
$\SGate\ket{a}=i^a\ket{a}$ (so $\ZGate=\SGate^2$) and the Hadamard gate (superposition) as $\HGate\ket{a}=\frac{1}{\sqrt{2}}\sum_b (-1)^{ab}\ket{b}$ (so $\XGate=\HGate\ZGate\HGate)$. 

We need at least one more quantum gate to obtain an approximately complete gate-set \cite{Dawson2005solovay_kitaev_thrm}. To this end we define the gate $\TGate\ket{a}=e^{i\pi/4}\ket{a}$, such that $S=T^2$. The Clifford+$\TGate$ gate-set plays an important role in our main result.

\Paragraph{Ancillas.}
To enable tradeoffs between circuit depth and circuit width, we permit the use of helper qubits, called ancillas, both clean (initialized to $\ket{0}$) and dirty (in an unknown state), as long as they are restored after use. 
Specifically, for a circuit $C$ over $n$ data qubits, $n_d$ dirty ancillas, and $n_c$ clean ancillas, we say that $C$ exactly implements the $n$-qubit unitary $U_C\in \Unitary{2^n}$ if for all $n$-qubit quantum states $\ket{\psi}$ and $n_d$-qubit quantum states $\ket{\chi}$, we have
$$C\ket{\psi}\ket{\chi}\ket{0^{n_c}} = (U_C\ket{\psi})\ket{\chi}\ket{0^{n_c}} \ .$$
We call such a circuit $C$ ancilla-restoring. 
A circuit that does not restore its ancillas implements no unitary by this definition; every circuit we construct is ancilla-restoring.
We identify a circuit with the unitary it exactly implements: we write $\Tr(C)$ for $\Tr(U_C)$ and speak of the eigenangles of $C$ for those of $U_C$.

\Paragraph{Special Unitaries.}
In the sequel, we will use three particular sets of unitaries:
\begin{align*}
    \SId &:= \{\IGate\} \ , \\
    \SCliff &:= \{U_C \colon C \text{ is a circuit over the Clifford gates } \HGate, \SGate, \CXGate\} \ .\\
    \GeneralPermutation & := \text{the set of generalized permutation matrices},
\end{align*}
where a generalized permutation matrix is a unitary of the form $\ket{x} \mapsto e^{ih(x)}\ket{L(x)}$ for a bijection $L$ on $\{0,1\}^n$ and a phase function $h\colon \{0,1\}^n\to \reals$.
Since they map basis states to basis states (with a global phase), they cannot introduce superposition.

Since the unitary $\IGate$ can be implemented by the empty circuit, $\SId\subseteq \SCliff$ and $\SId\subseteq \GeneralPermutation$
We will use the fact that $\SCliff$ is the normalizer of the Pauli group \cite[Ch.\ 10]{Nielsen_Chuang2010}. That is, if $U_C\in\SCliff$, then for any Pauli $P$ (a tensor of Pauli gates with phase $i^{\{0,1,2,3\}}$), we have $U_C P U_C^\dagger$ is a Pauli, and vice versa.

Note that the use of ancillas neither enlarges $\SCliff$ nor takes superposition-free circuits outside $\GeneralPermutation$.
Dirty ancillas reduce to clean ones: taking $\ket{\chi}=\ket{0^{n_d}}$ in the definition shows that $C$ also exactly implements $U_C$ when its $n_d$ dirty ancillas are regarded as clean, so we may assume $n_d=0$.
For superposition-free $C$, we have $C\colon\ket{x,y}\mapsto e^{ih(x,y)}\ket{L(x,y)}$, and restoration forces $L(x,0^{n_c})=(L'(x), 0^{n_c})$ for some bijection $L'$, so $U_C$ is again a generalized permutation matrix. 
For Clifford $C$, we use the normalizer property above. Let $P\in\mathcal{P}$ be any Pauli on the data qubits. Then
$C(P\otimes \IGate)C^\dagger\ket{\psi}\ket{0^{n_c}}=(U_C P U_C^\dagger \ket{\psi})\ket{0^{n_c}}$. 
By the normalizer property, $C(P\otimes \IGate)C^\dagger$ is also a Pauli, say $i^k(Q \otimes R)$; comparing the two expressions forces $R\ket{0^{n_c}}=i^\ell \ket{0^{n_c}}$ and $U_C P U_C^\dagger = i^{k-\ell}Q$, a Pauli. Hence $U_C$ is Clifford.

\subsection{Cost Functions and Free Sets} \label{subsec:cost_functions_and_free_sets}

We use a cost function to abstract from a particular implementation setting. A cost function is a map $f$ from circuits over $\gateset$ to non-negative reals. 
Our results depend on $f$ only through its free set $\GenericSet_f$, the set of unitaries that can be implemented at cost $0$:

$$\GenericSet_f := \{U_C \colon f(C) = 0\} \ ,$$

We consider every $f$ whose free set satisfies one of the following:
\begin{enumerate}
    \item $\GenericSet_f=\SId$.
    We call such $f$ a total cost function.
    Note that total gate depth and count are instances.
    \item $\IGate\in\GenericSet_f\subseteq\SCliff$.
    We call such $f$ a non-Clifford cost function.
    Note that $\TGate$-depth and $\TGate$-count are instances (when $\gateset$ is Clifford+$\TGate$).
    \item $\IGate\in\GenericSet_f\subseteq\GeneralPermutation$.
    We call such $f$ a superposition cost function; $\HGate$-depth and $\HGate$-count are instances (when $\gateset$ is Clifford+$\TGate$), as a circuit with no superposition-generating gate implements a generalized permutation matrix.
\end{enumerate}

\subsection{Distance Metrics} \label{sec:notions_of_distance}

The literature provides several functions for measuring the similarity between unitaries; we review three classes of them.
We say that a function $\Metric{U}{V}{m}$ from $n$-qubit unitaries $U,V\in \Unitary{2^n}$ to real numbers is a metric if these four properties hold: (i)~$\Metric{U}{V}{m} \geq 0$; (ii)~$\Metric{U}{V}{m}=\Metric{V}{U}{m}$; (iii)~$\Metric{U}{V}{m} \leq \Metric{U}{W}{m} + \Metric{W}{V}{m}$ for $n$-qubit $U,V,W\in \Unitary{2^n}$ (the triangle inequality); and (iv)~either~$\Metric{U}{V}{m}=0\iff U=V$ (which holds for the spectral metric), or~$\Metric{U}{V}{m}=0 \iff U \equiv V$ (which holds for the Sine and operator metrics).
If we have all but perhaps the triangle inequality, we simply call $m$ a distance measure.

We write $D_m$ for the diameter of $m$, that is, $D_m := \sup_{n}\sup_{U,V\in \Unitary{2^n}} \Metric{U}{V}{m}$.
We overload the notation $\Metric{\cdot}{\cdot}{m}$ so that either argument may be a circuit $C$ over $n$ data qubits, standing for the unitary $U_C$ it implements, and so that for a set of unitaries $\GenericSet$, we have $\Metric{\GenericSet}{U}{m} = \Metric{U}{\GenericSet}{m} := \inf_{V\in \GenericSet} \Metric{U}{V}{m}$, the infimum over distances between $U$ and elements of $\GenericSet$.

\Paragraph{Derived distance measures.}
We say that a distance measure $m_2$ is derived from $m_1$ if there is a strictly increasing $g\colon[0,D_{m_1}]\to\reals$, independent of $n$, s.t.\ $\Metric{U}{V}{m_2} = g(\Metric{U}{V}{m_1})$ for all $n$ and $U,V\in\Unitary{2^n}$.

\smallskip
\begin{lemma}
    \label{lemma:equivalence_of_metrics}
    Consider any gate-set $\gateset$.
    Let $\GenericSet$ be a set of unitaries, let $f$ be a cost function, let $m_2$ be a distance measure derived from $m_1$ via $g$, and let $0\leq\varepsilon\leq D_{m_1}$.
    Then $\ApproxMember(\GenericSet,m_1,\varepsilon,\gateset)=\ApproxMember(\GenericSet,m_2,g(\varepsilon),\gateset)$, and $\ApproxOpt(f,m_1,\varepsilon,\gateset)=\ApproxOpt(f,m_2,g(\varepsilon),\gateset)$,
    in the sense that they have the same yes-instances.
\end{lemma}

\begin{proof}
    As $g$ is strictly increasing, $\Metric{C}{U}{m_2}\leq g(\varepsilon)$ iff $\Metric{C}{U}{m_1}\leq\varepsilon$ for all $C$ and $U$.
\end{proof}

\Paragraph{Sine metric and derived trace measures.}
The Sine metric is defined in terms of the operator trace.
For a square matrix $A\in \Complex^{2^n\times 2^n}$ with eigenvalues $\{\lambda_j\}_j$, the trace is
$$\Tr(A):=\sum_{x\in \{0,1\}^n} \braket{x|A|x} = \sum_{j} \lambda_j \ .$$
We use the following standard properties: $\Tr(cA)=c \Tr(A)$, $\Tr(A\otimes B) = \Tr(A)\Tr(B)$, and $\Tr(\IGate)=2^n$ for the $n$-qubit identity $\IGate$.

The Sine metric~\cite{Gheorghiu2022TCountTDepth} is defined as 
$$\Metric{U}{V}{\SineMetric} := \sqrt{1-|\Tr(U^\dagger V)/2^n|^2}\ .$$
The literature contains many derived distance measures\footnote{These distance measures are often phrased in terms of the fidelity of two states. 
The fidelity evaluated on the Choi states of $U$ and $V$ equals $|\Tr(U^\dagger V)/2^n|^2$, so the state-level definitions and our trace-level definition describe the same quantity.}, including the Bures angle, Bures metric~\cite{Zhihao2009fidelity}, and the average gate distance~\cite{Nielsen2002}.
To illustrate, the Bures angle,
$\Metric{U}{V}{\BuresAngle}:= \cos^{-1}|\Tr(U^\dagger V)/2^n|$, is derived as $\Metric{U}{V}{\BuresAngle}=\sin^{-1}(\Metric{U}{V}{\SineMetric})$.
We have $D_\SineMetric = 1$.

\Paragraph{Spectral metric.}
The spectral metric is defined in terms of the operator norm.
We write $||\cdot||_2$ for the Euclidean vector norm.
Let $A,B\in \Complex^{2^n \times 2^n}$ be (not necessarily unitary) matrices.
The operator norm of $A$ is
$$\OperatorNorm{A}:=\sup_{||\vec{x}||_2=1} ||A\vec{x}||_2 \ .$$
If $A$ is normal, then $\OperatorNorm{A} = \max_{j} |\lambda_j|$, where $\{\lambda_j\}_j$ are the eigenvalues of $A$.
The operator norm satisfies $\OperatorNorm{A+B} \leq \OperatorNorm{A}+\OperatorNorm{B}$.
Further, $\OperatorNorm{e^{i\alpha}A} = \OperatorNorm{A}$ and $\OperatorNorm{A\otimes B} = \OperatorNorm{A}\OperatorNorm{B}$, while for unitary $U\in \Unitary{2^n}$, $\OperatorNorm{U}=1$.

For unitaries $U,V\in \Unitary{2^n}$ with spectral decomposition $U^\dagger V = QDQ^\dagger$ where $D=\diag(e^{i\phi_1},\dots,e^{i\phi_{2^n}})$, we have that $\OperatorNorm{U-V} \in [0,2]$, since
$$\OperatorNorm{U-V}=\OperatorNorm{\IGate-D} = \max_{j} |1-e^{i\phi_j}| = 2\max_j |\sin(\phi_j/2)| \enspace .$$

The spectral metric can now be defined as
$$\Metric{U}{V}{\SpectralAbbr} := \OperatorNorm{U-V} = 2 \max_j |\sin(\phi_j/2)| \ ,$$
where $\{\phi_j \}_j$ are the eigenangles of $U^\dagger V$.
Note that this metric is sensitive to global phase: 
there exist $U$, $V$, and $\alpha$ s.t.\ $\Metric{U}{V}{\SpectralAbbr} \neq \Metric{U}{e^{i\alpha}V}{\SpectralAbbr}$.
We have $D_\SpectralAbbr = 2$.

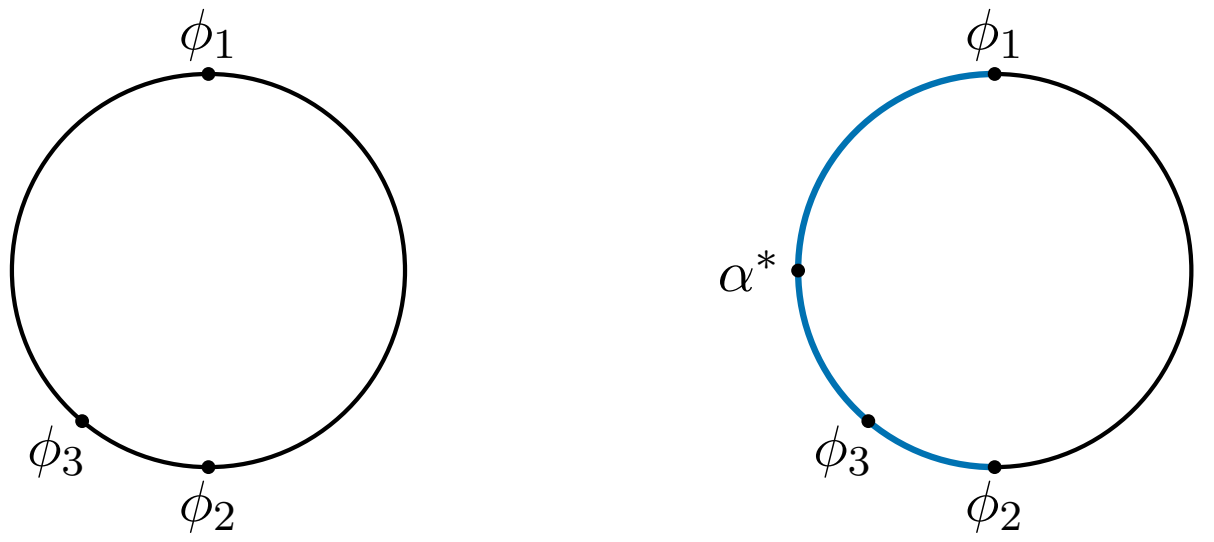
\begin{figure}[ht]
    \centering
    \begin{tikzpicture}[scale=1.3]

        \coordinate (C1) at (0,0);
        \coordinate (C3) at (4,0);
        
        \newcommand{\threecircle}[2]{%
            \draw[thick] (#1) circle (1);
            
            \foreach \angle/\name in {90/$\phi_1$, -90/$\phi_2$, 230/$\phi_3$} {
                \fill ($(#1)+(\angle:1)$) circle (1pt);
                \node at ($(#1)+(\angle:1.2)$) {\name};
            }
        }

        \newcommand{\circleline}[4]{%
            \draw[#4] ($(#1)+(#2:1)$) -- ($(#1)+(#3:1)$);
        }
        
        \threecircle{C1}{}

        \draw[thick] (C3) circle (1);
        
        \draw [otublue, very thick] (4,1) arc[start angle=90, end angle=270, radius=1];

        \foreach \angle/\name in {90/$\phi_1$, -90/$\phi_2$, 230/$\phi_3$} {
            \fill ($(4,0)+(\angle:1)$) circle (1pt);
            \node at ($(4,0)+(\angle:1.2)$) {\name};
        }

        \fill ($(4,0)+(180:1)$) circle (1pt);
        \node at ($(4,0)+(180:1.25)$) {$\alpha^*$};

    \end{tikzpicture}
    
    \caption{
        \emph{Left:}~A unit circle with the eigenangles of $U^\dagger V$.
        \emph{Right:}~The same unit circle with the shortest arc containing all eigenangles of $U^\dagger V$ drawn in. 
        The minimizing $\alpha^*$ sits at the midpoint of the arc.
    }
    \label{fig:eigenangles}
\end{figure}

\Paragraph{Operator metric.}
The operator metric is the global-phase-invariant counterpart of the spectral metric:   
$$\Metric{U}{V}{\OperatorAbbr}:= \min_{\alpha} \OperatorNorm{U-e^{i\alpha} V} = 2 \min_{\alpha} \max_j \Big|\sin\Big(\frac{\phi_j+\alpha}{2}\Big)\Big|\ ,$$
where $\{\phi_j\}_j$ are the eigenangles of $U^\dagger V$.
Using standard trigonometry, as the optimizing $\alpha^*$ minimizes the worst case, it lies at the midpoint of a shortest arc on the unit circle containing all eigenangles, meaning the endpoints of the arc are the worst case (see~\cref{fig:eigenangles}).
We see that an equivalent formula is
$$\Metric{U}{V}{\OperatorAbbr} = 2 \sin (\Delta/4)\ ,$$
where $\Delta$ is the length of the shortest arc on the unit circle containing all eigenangles of $U^\dagger V$.
We have $D_\OperatorAbbr = 2$.

\Paragraph{Mutually underivable metrics.}
No two of the Sine, spectral, and operator metrics are derived from one another, so~\cref{lemma:equivalence_of_metrics} does not transfer results between them and we treat each separately.
Specifically, for any two distinct metrics $m_1, m_2$ among the Sine, spectral, and operator metrics, there exist unitaries $A$ and $B$, s.t.\ $\Metric{A}{I}{m_1} = \Metric{B}{I}{m_1}$ but $\Metric{A}{I}{m_2} \neq \Metric{B}{I}{m_2}$.
For spectral vs.\ operator, we have $\Metric{\IGate}{\IGate}{\OperatorAbbr} = \Metric{\IGate}{-\IGate}{\OperatorAbbr} = 0$, while $\Metric{\IGate}{\IGate}{\SpectralAbbr} = 0 \neq 2 = \Metric{\IGate}{-\IGate}{\SpectralAbbr}$. 
For Sine vs.\ spectral, and Sine vs.\ operator, one can verify that $\Metric{\IGate\otimes \IGate}{\CZGate}{\SpectralAbbr}=\Metric{\IGate \otimes \IGate}{\IGate\otimes \ZGate}{\SpectralAbbr}$ and $\Metric{\IGate\otimes \IGate}{\CZGate}{\OperatorAbbr}=\Metric{\IGate \otimes \IGate}{\IGate\otimes \ZGate}{\OperatorAbbr}$, while $\Metric{\IGate\otimes \IGate}{\CZGate}{\SineMetric}\neq\Metric{\IGate \otimes \IGate}{\IGate\otimes \ZGate}{\SineMetric}$.

\subsection{The complexity class $\PP$}
Let $\varphi\colon \{0,1\}^n \to \{0,1\}$ be a propositional formula over $n$ bits, defined inductively over conjunctions, disjunctions, and negations of Boolean variables.
We write $\FRACSAT{\varphi} := |\{x\colon \varphi(x)\}|/2^n$ for the fraction of satisfying assignments of $\varphi$.
Note that $\FRACSAT{\varphi}+\FRACSAT{\neg\varphi}=1$.

The complexity class $\PP$~\cite{gill1974computational} is central to our results.
A language $L$ is in $\PP$ if there is a probabilistic polynomial-time Turing machine that accepts $x$ with probability greater than $1/2$ iff $x\in L$.
A real number $r$ is $\PTIME$-time computable if, given $b$ in unary, a dyadic rational within $2^{-b}$ of $r$ can be computed in time $b^{O(1)}$.
The $\PTIME$-time computable reals remain $\PTIME$-computable under $r\mapsto\sqrt{r}$ on $[0,\infty)$, $r\mapsto\sin r$, and $r\mapsto\sin^{-1} r$ on $(-1,1)$~\cite{brent1976fast,ko1991complexity}.
It is folklore that $\ATMOSTSAT(\beta) := \{\varphi\colon \FRACSAT{\varphi} \leq \beta\}$ is $\PP$-hard for every fixed $\PTIME$-time computable $0<\beta<1$; for $\beta=1/2$, see~\cite{gill1974computational,simon1975some}.

\Paragraph{Karp reductions.}
A Karp reduction from a decision problem $\Problem{P}$ to a decision problem $\Problem{Q}$ is a $\PTIME$-time computable map $r$ s.t.\ $x$ is a yes-instance of $\Problem{P}$ iff $r(x)$ is a yes-instance of $\Problem{Q}$; if $\Problem{P}$ is a promise problem, this is required only for $x$ satisfying the promise of $\Problem{P}$, and if $\Problem{Q}$ is a promise problem, $r(x)$ must satisfy its promise.
A problem is $\PP$-hard if every problem in $\PP$ Karp-reduces to it, so $\PP$-hardness transfers along Karp reductions.

Since $\PP$ contains $\BQP$, the class of ``quantum tractable'' problems, an efficient classical algorithm for any $\PP$-hard problem would undermine the computational advantage of quantum computing.
Moreover, $\PP$ is not even contained in the Polynomial Hierarchy unless the Polynomial Hierarchy collapses~\cite{Toda91}, placing it well beyond what is considered tractable.
Further, $\PP$ contains $\NQP$ and is itself contained in $\PSPACE$.

\section{The \texorpdfstring{$\PP$}{PP}-hardness of Approximate Identity} \label{sec:framework_for_pp_hardness}

In this section, we prove the case $\GenericSet=\SId$ of \cref{theorem:approx_member_hardness}, i.e., of checking approximate identity.
We prove the stronger statement that a promise version of $\ApproxIdentity$ is $\PP$-hard.
The hardness of the promise version will allow us to also establish hardness for $\ApproxMember(\GenericSet)$ with $\IGate\in\GenericSet\subseteq\SCliff$ in the next section (cf.\ \cref{lemma:reduction_promise_approx_id_to_approx_member}).

\subsection{Promise Problems}\label{subsec:promise_problems}

We begin with a promise version of $\ATMOSTSAT$, parameterized by real numbers $\delta$ and $\kappa$.

\titlebox{$\PROMISEATMOSTSAT(\delta, \kappa)$}{
\emph{Input:} An $n$-bit propositional formula $\varphi$. \\
\emph{Promise:} $\FRACSAT{\varphi} \leq \kappa$. \\
\emph{Question:} Is $\FRACSAT{\varphi} \leq \delta$?
}

The next lemma states that $\PROMISEATMOSTSAT$ is also $\PP$-hard.
The proof is via a reduction from $\ATMOSTSAT(\beta)$ that scales the satisfying fraction by a fixed dyadic constant; although it appears to be folklore, we are not aware of a specific reference for this fact, and thus we present it in the appendix.

\smallskip

\begin{lemma}
    \label{lemma:hardness_promise_atmostsat}
    Let $0<\delta<\kappa\leq1$ be fixed, with $\delta$ $\PTIME$-time computable.
    Then $\PROMISEATMOSTSAT(\delta, \kappa)$ is $\PP$-hard.
\end{lemma}

Next, we define a promise version of $\ApproxMember$, where an additional real number $\tau$ serves as an approximation promise.

\titlebox{$\PromiseApproxMember(\GenericSet,m,\varepsilon,\tau,\gateset)$}{
\emph{Input:} A circuit $C$ over $\gateset$. \\
\emph{Promise:} $\Metric{\GenericSet}{C}{m} \leq \tau$. \\
\emph{Question:} Is $\Metric{\GenericSet}{C}{m} \leq \varepsilon$?
}
We further let $\PromiseApproxIdentity(m,\varepsilon,\tau,\gateset)$ be $\PromiseApproxMember(\SId,m,\varepsilon,\tau,\gateset)$.
The next lemma states sufficient conditions on $\GenericSet$, $\gateset$, and $m$ that make $\PromiseApproxMember$ hard.
In \cref{sec:lower_bounds_on_approx_ID}, we establish the hardness of $\PromiseApproxIdentity$ under the Sine, spectral, and operator metrics by showing that it meets the conditions of the lemma.

\smallskip
\begin{lemma}
    \label{lemma:template}
    Consider any gate-set $\gateset$ that exactly implements Clifford+$\TGate$.
    Fix a set of unitaries $\GenericSet$, a metric $m$, and a constant $0<\kappa\leq 1$.
    Let $g\colon [0,1]\to\reals$ be strictly increasing and continuous on $[0,\kappa]$ with $g(0)=0$.
    Suppose there is a $\PTIME$-time computable mapping from propositional formulas $\varphi$ with $\FRACSAT{\varphi}\leq \kappa$ to ancilla-restoring Clifford+$\TGate$ circuits $C$ s.t.\ 
    $$g(\FRACSAT{\varphi})=\Metric{\GenericSet}{C}{m}\enspace .$$ 
    Then $\PromiseApproxMember(\GenericSet,m,\varepsilon,g(\kappa),\gateset)$ is $\PP$-hard for all fixed $0<\varepsilon<g(\kappa)$ s.t.\ $g^{-1}(\varepsilon)$ is $\PTIME$-time computable, where $g^{-1}$ is the inverse of $g$ on $[0,\kappa]$.
\end{lemma}

\begin{proof}
    Let $0<\varepsilon<g(\kappa)$ be given s.t.\ $g^{-1}(\varepsilon)$ is $\PTIME$-time computable, where $g^{-1}$ exists on $[0,\kappa]$ since $g$ is strictly increasing and continuous there. Set $\delta:=g^{-1}(\varepsilon)$, which is $\PTIME$-time computable. As $g(0)=0$, we have $0<\delta<\kappa$.
    We give a Karp reduction from $\PROMISEATMOSTSAT(\delta,\kappa)$ to $\PromiseApproxMember(\GenericSet,m,\varepsilon,g(\kappa),\gateset)$, which yields the result by~\cref{lemma:hardness_promise_atmostsat}.

    Given a yes/no instance $\varphi$ of $\PROMISEATMOSTSAT(\delta,\kappa)$, transform $\varphi$ into a Clifford+$\TGate$ circuit $C$ using the $\PTIME$-time computable mapping.
    Since $\gateset$ exactly implements Clifford+$\TGate$, replacing each gate of $C$ by its implementation over $\gateset$ yields a circuit $C'$ over $\gateset$ with $U_{C'} = U_C$, which is the output of the reduction.

    We proceed to show correctness, that is, $\varphi$ is a yes-instance iff $C'$ is a yes-instance.
    By the promise of $\PROMISEATMOSTSAT(\delta,\kappa)$, we have $0\leq \FRACSAT{\varphi} \leq \kappa$, so
    $$g(\FRACSAT{\varphi}) = \Metric{\GenericSet}{C}{m} = \Metric{\GenericSet}{C'}{m} \ ,$$
    where the second equality follows from $U_C = U_{C'}$.
    Hence $C'$ is a yes-instance iff $g(\FRACSAT{\varphi})\leq\varepsilon$.
    As $g$ is strictly increasing on $[0,\kappa]$, this is equivalent to $\FRACSAT{\varphi}\leq g^{-1}(\varepsilon)=\delta$, which is exactly the acceptance condition of $\PROMISEATMOSTSAT(\delta,\kappa)$.

    It remains to verify the promise, that is, $\Metric{\GenericSet}{C'}{m} \leq g(\kappa)$.
    By the promise of $\PROMISEATMOSTSAT(\delta,\kappa)$, we have $\FRACSAT{\varphi} \leq \kappa$, and since $g$ is increasing on $[0,\kappa]$, we see that $\Metric{\GenericSet}{C'}{m} = g(\FRACSAT{\varphi})\leq g(\kappa)$ as required.
\end{proof}

\subsection{Hardness of \texorpdfstring{$\PromiseApproxIdentity$}{PromiseApproxIdentity}} \label{sec:lower_bounds_on_approx_ID}

In this subsection, we apply~\cref{lemma:template} to obtain hardness of $\PromiseApproxIdentity$. 
For each metric, we provide a $\PTIME$-time computable mapping from propositional formulas $\varphi$ to circuits $C$, and a closed form for $\Metric{\SId}{C}{m}=\Metric{\IGate}{C}{m}$ in terms of $\FRACSAT{\varphi}$. 
In the rest of this section, we prove the following theorem.

\smallskip
\begin{theorem}
    \label{theorem:approx_identity_promise_hardness}
    Consider any gate-set $\gateset$ that exactly implements Clifford+$\TGate$, and let $m$ be the Sine, spectral, or operator metric.
    Then $\PromiseApproxIdentity(m,\varepsilon,\tau,\gateset)$ is $\PP$-hard for all $0 < \varepsilon < \tau < D_m$ with $\varepsilon$ $\PTIME$-time computable.
\end{theorem}

We treat the Sine, spectral, and operator metrics in turn, in each case applying~\cref{lemma:template}. The spectral metric additionally depends on~\cref{lemma:spectral_construction} and the operator metric additionally on~\cref{lemma:op_construction}.

\subsubsection{Proof of \cref{theorem:approx_identity_promise_hardness} for the Sine Metric}
Let $0 < \tau < 1$ be given.
Given $\varphi$ over $n$ bits, we construct a Clifford+$\TGate$ circuit $C$ over $n+1$ data qubits with $U_C \colon \ket{x}\ket{b} \mapsto \ket{x}\ket{b \oplus \varphi(x)}$.
Treat $\varphi$ as a classical Boolean circuit in the obvious way, and apply the construction of Section 1.4.1 of Nielsen and Chuang~\cite{Nielsen_Chuang2010} to obtain an $\XGate$+$\CXGate$+$\TOF$ circuit with this unitary that uses polynomially many clean ancillas.
Replacing each $\XGate$ and $\TOF$ gate by its implementation over Clifford+$\TGate$ yields $C$.
Set $g(x) := \sqrt{1 - (1 - x)^2}$ and $\kappa := g^{-1}(\tau) \in (0, 1)$. 
Then $g(0) = 0$, $g(\kappa) = \tau > 0$, and $g$ is strictly increasing and continuous on $[0, \kappa]$.
To apply~\cref{lemma:template}, it suffices to show that for $0\leq \FRACSAT{\varphi}\leq \kappa$,
$$g(\FRACSAT{\varphi})=\Metric{\SId}{C}{\SineMetric} = \Metric{\IGate}{C}{\SineMetric} \ .$$ 
Towards this, we compute
\begin{align*}
    \Tr(C)
        &= \sum_{x,b} \braket{x,b|C|x,b} \\
        &= \sum_{x,b} \braket{x,b|x,b\oplus \varphi(x)} \\
        &= \sum_{x,b} \braket{b|b\oplus \varphi(x)} \\
        &= 2 \sum_{x \colon \neg \varphi(x)} 1 \\
        &= 2^{n+1} \FRACSAT{\neg \varphi} \ .
\end{align*}
Plugging this in, we see that 
\begin{align*}
    \Metric{\IGate}{C}{\SineMetric} 
        &= \sqrt{1-|\Tr(C) / 2^{n+1}|^2} \\
        &= \sqrt{1-|\FRACSAT{\neg \varphi}|^2} \\
        &= \sqrt{1-\FRACSAT{\neg \varphi}^2} \\
        &= \sqrt{1-(1-\FRACSAT{\varphi})^2} \\
        &= g(\FRACSAT{\varphi}) \ .
\end{align*}
Hence $g(\FRACSAT{\varphi}) = \Metric{\IGate}{C}{\SineMetric}$.
For $\PTIME$-time computable $0<\varepsilon<\tau$, the value $g^{-1}(\varepsilon) = 1-\sqrt{1-\varepsilon^2}$ is $\PTIME$-time computable, so~\cref{lemma:template} yields $\PP$-hardness of $\PromiseApproxIdentity(\SineMetric,\varepsilon,\tau,\gateset)$ for every $\PTIME$-time computable $0 < \varepsilon < \tau<1$. This concludes the proof for the Sine metric.

\smallskip
\begin{remark}
    \label{rem:sine_hardness_classical_reversible}
	Before the final substitution, the circuit constructed above for the Sine metric is a standard oracle for $\varphi$ that uses only $\XGate$, $\CXGate$, and $\TOF$ gates.
	Hence, for any gate-set $\gateset$ that exactly implements $\XGate$, $\CXGate$, and $\TOF$, $\ApproxIdentity(\SineMetric, \varepsilon, \gateset)$ is $\PP$-hard for every $\PTIME$-time computable $0<\varepsilon < 1$, even when $\gateset$ has no access to superposition or phase (i.e., is classical-reversible).
\end{remark}

\subsubsection{Proof of \cref{theorem:approx_identity_promise_hardness} for the Spectral Metric}

The spectral and operator metrics are functions of the spectrum, so to apply~\cref{lemma:template} to these metrics, we construct a unitary whose spectrum is a function of the fraction of satisfying assignments of $\varphi$, namely of
$$\theta(\varphi) := 2 \sin^{-1}\sqrt{\FRACSAT{\varphi}} \ .$$
As $x \mapsto 2\sin^{-1}\sqrt{x}$ is a strictly increasing continuous bijection from $[0,1]$ onto $[0,\pi]$ with inverse $y \mapsto \sin^2(y/2)$, we have $0 \leq \theta(\varphi) \leq \pi$, $\FRACSAT{\varphi} = \sin^2(\theta(\varphi)/2)$, and $\theta(\varphi) \leq \theta_0$ iff $\FRACSAT{\varphi} \leq \sin^2(\theta_0/2)$ for every $\theta_0 \in [0,\pi]$.
The following lemma gives such a unitary.
Replacing $\varphi$ by $\varphi'(x,y) := \varphi(x)$ for a fresh bit $y$ preserves $\FRACSAT{\varphi}$ and takes linear time, so we assume $n \geq 2$ in what follows.
For a satisfiable propositional formula $\varphi$ over $n$ bits, we define
$$\ket{\varphi}:= \frac{1}{\sqrt{2^n \FRACSAT{\varphi}}} \sum_{x\colon \varphi(x)} \ket{x} \ .$$

\smallskip
\begin{lemma}
    \label{lemma:spectral_construction}
    Given a propositional formula $\varphi$ over $n \geq 2$ bits, there is a $\PTIME$-time constructible Clifford+$\TGate$ circuit $C$ s.t.\ the eigenangles of $C$ are $\{0,\pm 2\theta\}$, where $\theta= \theta(\varphi)$.
    Moreover, if $\varphi$ and $\neg\varphi$ are both satisfiable, then
    $$U_C = \IGate - 2\sin^2(\theta)\big(\ket{\neg\varphi}\bra{\neg\varphi} + \ket{\varphi}\bra{\varphi}\big)
            + \sin (2\theta)\big(\ket{\varphi}\bra{\neg\varphi} - \ket{\neg\varphi}\bra{\varphi}\big) \ ,$$
    and otherwise $U_C = \IGate$.
\end{lemma}

\begin{proof}
    We take $C:= G^2$, where $G$ is the Grover iterate~\cite{Grover1996FastQuantumMechanicalAlgorithm} of $\varphi$:
    $$G := (2\ket{+^n}\bra{+^n}-\IGate)(\IGate - 2\sum_{x\colon \varphi(x)} \ket{x}\bra{x})\enspace ,$$
    where the left factor is Grover's diffusion operator (reflection about $\ket{+^n}$) and the right factor is the phase oracle for $\varphi$ (mapping $\ket{x}$ to $(-1)^{\varphi(x)}\ket{x}$).
    Since $G$ is implemented by a polynomially sized, $\PTIME$-time constructible ancilla-restoring Clifford+$\TGate$ circuit (Figure 6.2 of~\cite{Nielsen_Chuang2010}), so is $C$.

    If $\varphi$ is unsatisfiable, then $G = 2\ket{+^n}\bra{+^n}-\IGate$, and if $\neg\varphi$ is unsatisfiable, then $G = \IGate - 2\ket{+^n}\bra{+^n}$. In either case, $G$ is a reflection, so $U_C = G^2 = \IGate$,
    and $\theta\in\{0,\pi\}$, so the eigenangles of $C$ are $\{0\} = \{0,\pm 2\theta\} \pmod{2\pi}$.
    
    Otherwise $\ket{\varphi}$ and $\ket{\neg\varphi}$ are well-defined states; let $\Pi := \ket{\neg\varphi}\bra{\neg\varphi} + \ket{\varphi}\bra{\varphi}$ be the projector onto the span of $\ket{\varphi}$ and $\ket{\neg\varphi}$.
    The state $\ket{+^n} = \cos(\theta/2)\ket{\neg\varphi} + \sin(\theta/2)\ket{\varphi}$ lies in the span of $\ket{\varphi}$ and $\ket{\neg\varphi}$, and $G$ restricted to this span is the rotation $G\ket{\neg\varphi} = \cos\theta\ket{\neg\varphi} + \sin\theta\ket{\varphi}$, $G\ket{\varphi} = -\sin\theta\ket{\neg\varphi} + \cos\theta\ket{\varphi}$ (Section~6.1.3 of~\cite{Nielsen_Chuang2010}).
    Hence $U_C\ket{\neg\varphi} = \cos 2\theta\ket{\neg\varphi} + \sin 2\theta\ket{\varphi}$ and $U_C\ket{\varphi} = -\sin 2\theta\ket{\neg\varphi} + \cos 2\theta\ket{\varphi}$.
    For $v$ perpendicular to the span of $\ket{\varphi}$ and $\ket{\neg\varphi}$, write $v = v_1 + v_0$ with $v_1$ supported on $\{x\colon\varphi(x)\}$ and $v_0$ on $\{x\colon\neg\varphi(x)\}$.
    Then $\braket{\varphi|v_1} = \braket{\varphi|v} = 0$ and $\braket{\neg\varphi|v_0} = \braket{\neg\varphi|v} = 0$, so $v_0$ and $v_1$ are also perpendicular to this span.
    The reflection $\IGate - 2\sum_{x\colon \varphi(x)} \ket{x}\bra{x}$ maps $v$ to $v_0 - v_1$, which is perpendicular to this span and hence to $\ket{+^n}$, so $Gv = v_1 - v_0$ and $U_C v = G(v_1 - v_0) = v$.
    Combining both, we obtain
    \begin{align*}
        U_C
            &= U_C(\IGate-\Pi) + U_C\Pi \\
            &= (\IGate-\Pi) + (U_C\ket{\neg\varphi})\bra{\neg\varphi} + (U_C\ket{\varphi})\bra{\varphi}
                & \left[U_C v = v \text{ for } v \perp \ket{\varphi},\ket{\neg\varphi}\right] \\
            &= (\IGate - \Pi) + \cos(2\theta)\Pi + \sin(2\theta)\big(\ket{\varphi}\bra{\neg\varphi} - \ket{\neg\varphi}\bra{\varphi}\big) \\
            &= \IGate - 2\sin^2(\theta)\Pi + \sin(2\theta)\big(\ket{\varphi}\bra{\neg\varphi} - \ket{\neg\varphi}\bra{\varphi}\big)
                & \left[1-\cos 2\theta = 2\sin^2\theta\right] \ .
    \end{align*}
    In particular, $U_C$ is the rotation by $2\theta$ on the span of $\ket{\varphi}$ and $\ket{\neg\varphi}$, with eigenangles $\pm 2\theta$, and the identity on its orthogonal complement, which has dimension $2^n-2\geq 2$, with eigenangle $0$,
    so the eigenangles of $C$ are $\{0,\pm 2\theta\}$.
\end{proof}

Let $0 < \tau < 2$ be given. 
Given $\varphi$, construct $C$ as in~\cref{lemma:spectral_construction}, so that $C$ has eigenangles $\{0, \pm2\theta\}$ with $\theta = \theta(\varphi)$. 
Set $g(x) := 4\sqrt{x(1-x)}$ and $\kappa := g^{-1}(\tau)=(1 - \sqrt{1-\tau^2/4})/2 \in (0, 1/2)$.
Here $g$ increases on $[0, 1/2]$ and decreases on $[1/2, 1]$, with $g(1/2) = 2 = D_\SpectralAbbr$, so we read $g^{-1}$ as the inverse of the restriction of $g$ to $[0, 1/2]$, on which it is a continuous bijection onto $[0, 2]$; since $\tau < 2$ this gives $\kappa < 1/2$.
Then $g(0)=0$, $g(\kappa)=\tau>0$, and $g$ is strictly increasing and continuous on $[0,\kappa]$.
To apply~\cref{lemma:template}, it suffices to show that $g(\FRACSAT{\varphi}) = \Metric{\SId}{C}{\SpectralAbbr} = \Metric{\IGate}{C}{\SpectralAbbr}$ for $0 \leq \FRACSAT{\varphi} \leq \kappa$.
We have
\begin{align*}
    \Metric{\IGate}{C}{\SpectralAbbr} 
        &= 2\max \{ |\sin(0/2)|, |\sin(\pm 2\theta/2)| \} \\
        &= 2\max \{0,|\sin\theta|\} & \left[|\sin (-\alpha)| = |\sin \alpha|\right] \\
        &= 2|\sin \theta| \ .
\end{align*}
We see that
\begin{align*}
    2|\sin \theta| 
        &= 2|\sin \theta(\varphi) | \\
        &= 2\big| \sin \big(2\sin^{-1}\sqrt{\FRACSAT{\varphi}}\big)\big| \\
        &= 4\big|\sin \big(\sin^{-1}\sqrt{\FRACSAT{\varphi}} \big) \cos \big(\sin^{-1}\sqrt{\FRACSAT{\varphi}} \big)\big| & \left[\sin 2x = 2\sin x\cos x\right] \\
        &= 4\big| \sqrt{\FRACSAT{\varphi}} \sqrt{1-\FRACSAT{\varphi}}\big| & \left[\cos(\sin^{-1} y) = \sqrt{1-y^2}\right] \\
        &= 4 \sqrt{\FRACSAT{\varphi}} \sqrt{1-\FRACSAT{\varphi}} & \left[\sqrt{\FRACSAT{\varphi}},\sqrt{1-\FRACSAT{\varphi}} \geq 0\right] \\
        &= 4 \sqrt{\FRACSAT{\varphi}(1-\FRACSAT{\varphi})} \\
        &= g(\FRACSAT{\varphi}) \ .
\end{align*}
For $\PTIME$-time computable $0<\varepsilon<\tau$, the value $g^{-1}(\varepsilon) = (1-\sqrt{1-\varepsilon^2/4})/2$ is $\PTIME$-time computable, so~\cref{lemma:template} yields $\PP$-hardness of $\PromiseApproxIdentity(\SpectralAbbr,\varepsilon,\tau,\gateset)$ for every $\PTIME$-time computable $0 < \varepsilon < \tau < 2$. This concludes the proof for the spectral metric.

\subsubsection{Proof of \cref{theorem:approx_identity_promise_hardness} for the Operator Metric}
In the remainder of this section, let $\ell>0$ be a constant integer (depending on $\tau$), to be chosen later, and set
$$\theta_{\max} := \frac{\pi}{2^{\ell+1}-1} \ , \quad \gamma_{\max} := \sin^2 (\theta_{\max}/2) \ , \quad \Theta_\ell := (2^\ell - 1)\theta_{\max} \ .$$
The operator metric is a function of the length of the shortest arc on the unit circle containing all eigenangles, so to apply~\cref{lemma:template}, we control the length of this arc, which means controlling the relative distances between all eigenangles.
The following lemma controls this arc by taking a tensor product of powers of the circuit of~\cref{lemma:spectral_construction}.

\smallskip
\begin{lemma}
    \label{lemma:op_construction}
    Given a propositional formula $\varphi$ with $0\leq \FRACSAT{\varphi}\leq \gamma_{\max}$, there is a $\PTIME$-time constructible Clifford+$\TGate$ circuit $C$ s.t.
    $$\Metric{\IGate}{C}{\OperatorAbbr} = 2\sin ((2^\ell - 1)\theta(\varphi)) \ .$$
\end{lemma}

\begin{figure}[ht]
    \centering
    \begin{tikzpicture}[scale=1.6]
        \coordinate (C) at (0,0);

        \draw[thick] (C) circle (1);
        \draw[otublue, very thick] ($(C)+(-140:1)$)
            arc[start angle=-140, end angle=140, radius=1];

        \foreach \a in {-140,-120,...,140} {
            \fill ($(C)+(\a:1)$) circle (1pt);
        }

        \draw[<->] ($(C)+(0:0.86)$)
            arc[start angle=0, end angle=20, radius=0.86];
        \node at ($(C)+(10:0.70)$) {$2\theta$};

        \node at ($(C)+(0:1.6)$) {$\Delta$};

        \node at ($(C)+(180:1.6)$) {$2\pi-\Delta$};

        \node at ($(C)+(140:1.46)$) {$2(2^\ell-1)\theta$};
        \node at ($(C)+(-140:1.46)$) {$-2(2^\ell-1)\theta$};
    \end{tikzpicture}

    \caption{
        The eigenangles of $C$ in the proof of~\cref{lemma:op_construction},
        drawn for $\ell=3$: the multiples of $2\theta$ from
        $-2(2^\ell-1)\theta$ to $2(2^\ell-1)\theta$.
        The highlighted arc contains all eigenangles and has length $\Delta=4(2^\ell-1)\theta$.
        The complement of this arc is a gap of length $2\pi-\Delta$.        
        As long as the gap is at least as long as the spacing between consecutive eigenangles, meaning
        $2\pi-\Delta\ge 2\theta$ (or equivalently, $\theta\leq \theta_{\max}$), the shortest arc containing all eigenangles is
        the highlighted one, and $\Metric{\IGate}{C}{\OperatorAbbr}=2\sin(\Delta/4)$.
    }
    \label{fig:eigenangle-placement}
\end{figure}
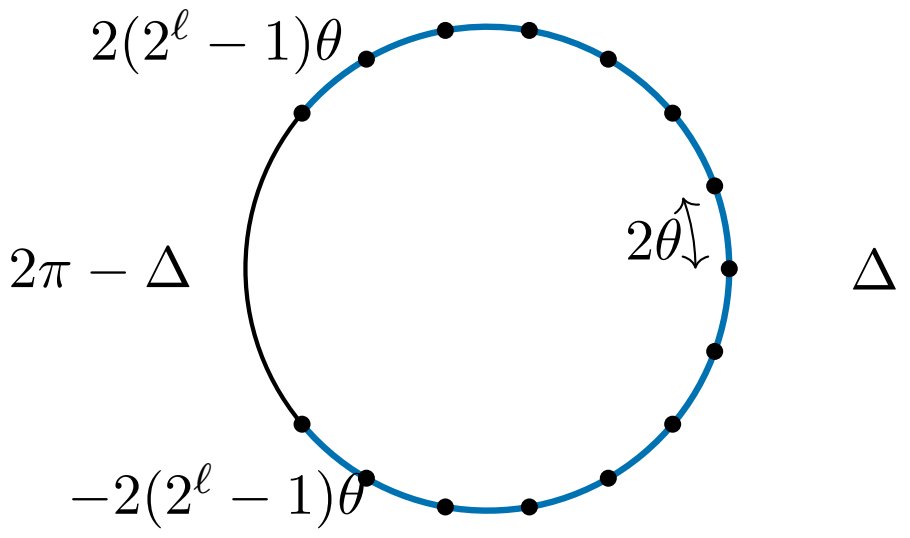

\begin{proof}
    Let $\varphi$ be given.
    Note that $0\leq \FRACSAT{\varphi}\leq \gamma_{\max}$ iff $\theta(\varphi) \leq \theta_{\max}$.
    Let $C_1$ be the Clifford+$\TGate$ circuit obtained by applying~\cref{lemma:spectral_construction} to $\varphi$.
    By~\cref{lemma:spectral_construction}, $C_1$ has eigenangles $\{0,\pm 2\theta\}$, where $\theta=\theta(\varphi)$.
    Then let 
    $$C := C_1^{2^0} \otimes C_1^{2^1} \otimes C_1^{2^2} \otimes \dots \otimes C_1^{2^{\ell-1}} \ .$$
    Note that $C$ is the tensor product of Clifford+$\TGate$ circuits and hence is a Clifford+$\TGate$ circuit. 
    Since $\ell$ is constant, $2^j$ for $0\leq j \leq \ell-1$ is constant, and since $C_1$ is polynomially sized, $C_1^{2^{j}}$ is also polynomially sized.
    As $C$ is the tensor product of $\ell$ such circuits, then $C$ is also polynomially sized and hence $\PTIME$-time constructible.

    The eigenangles of $C_1^{2^j}$ are $\{0,\pm 2\theta \cdot 2^j\}$, so the eigenangles of $C$ are
    $$\left\{ \sum_{j=0}^{\ell-1} a_j \colon a_j \in \{0,\pm 2\theta \cdot 2^j\} \right\} = \left\{ \pm 2\theta \cdot \sum_{j=0}^{\ell-1} 2^j b_j \right\} \ ,$$
    where $b_0,b_1, \dots, b_{\ell-1}$ are bits.
    These eigenangles are multiples of $2\theta$ from $-2(2^\ell - 1)\theta$ to $2(2^\ell - 1)\theta$.
    See~\cref{fig:eigenangle-placement} for their placement on the unit circle.
    The eigenangles lie on an arc of length $\Delta = 4(2^\ell-1)\theta$, whose complement has length $2\pi-\Delta$.
    If $2\pi-\Delta \geq 2\theta$, the complement is the largest gap between consecutive eigenangles, so the arc of length $\Delta$ is the shortest arc containing all eigenangles. 
    The condition $2\pi - \Delta \geq 2\theta$ rearranges to $\theta \leq \pi/(2^{\ell+1} - 1) = \theta_{\max}$, which, as noted at the start of the proof, is equivalent to $\FRACSAT{\varphi} \leq \gamma_{\max}$. 
    In that range,
    $$\Metric{\IGate}{C}{\OperatorAbbr} = 2\sin(\Delta/4) = 2\sin((2^\ell-1)\theta) \ .$$
\end{proof}

Let $0 < \tau < 2$ be given.
By definition of $\Theta_\ell$ and $\theta_{\max}$,
$$\Theta_\ell = \frac{\pi(2^\ell-1)}{2^{\ell+1}-1}\ ,$$
so $\Theta_\ell<\pi/2$ and $\Theta_\ell\to\pi/2$ as $\ell\to\infty$; hence $\sup_\ell 2\sin\Theta_\ell = 2$.
We now fix $\ell>0$ with $2 \sin \Theta_\ell > \tau$.
Given $\varphi$, construct $C$ as in~\cref{lemma:op_construction} for this $\ell$, and set $g(x) := 2 \sin\big(2(2^\ell - 1)\sin^{-1}\sqrt{x}\big)$, so that $g(\FRACSAT{\varphi}) = 2\sin((2^\ell - 1)\theta(\varphi))$.
Since $x \mapsto 2\sin^{-1}\sqrt{x}$ is strictly increasing and continuous, maps $0$ to $0$, and maps $\gamma_{\max}$ to $\theta_{\max}$, the outer sine in $g$ is applied only on $[0, \Theta_\ell] \subseteq [0, \pi/2)$ for $x \in [0, \gamma_{\max}]$, where it is strictly increasing and continuous.
Hence $g$ is strictly increasing and continuous on $[0, \gamma_{\max}]$ with $g(0) = 0$ and $g(\gamma_{\max}) = 2 \sin \Theta_\ell > \tau$, so $g$ is invertible there and $\kappa := g^{-1}(\tau)$ satisfies $\kappa < \gamma_{\max}$.
Therefore the closed form of~\cref{lemma:op_construction} applies throughout $[0, \kappa]$, where it reads $g(\FRACSAT{\varphi}) = \Metric{\SId}{C}{\OperatorAbbr} = \Metric{\IGate}{C}{\OperatorAbbr}$, and $g(\kappa) = \tau > 0$ with $g$ strictly increasing and continuous on $[0, \kappa]$. 
For $\PTIME$-time computable $0<\varepsilon<\tau$, the value $g^{-1}(\varepsilon) = \sin^2\big(\sin^{-1}(\varepsilon/2)/(2(2^\ell-1))\big)$ is $\PTIME$-time computable, so~\cref{lemma:template} yields $\PP$-hardness of $\PromiseApproxIdentity(\OperatorAbbr, \varepsilon, \tau, \gateset)$ for every $\PTIME$-time computable $0 < \varepsilon < \tau<2$.
This completes the proof of~\cref{theorem:approx_identity_promise_hardness}.

\section{The \texorpdfstring{$\PP$}{PP}-hardness of Approximate Membership} \label{sec:approx_member_instances_hardness}

In this section, we turn our attention to the hardness of $\ApproxMember$ over sets $\GenericSet$ with $\IGate\in\GenericSet\subseteq \SCliff$ or $\IGate\in\GenericSet\subseteq \GeneralPermutation$, and prove~\cref{theorem:approx_member_hardness} and~\cref{corollary:approx_opt_hardness}.
\cref{subsec:id_to_general_free_sets} treats $\GenericSet\subseteq\SCliff$ by a reduction from $\PromiseApproxIdentity$ (\cref{theorem:hardness_approx_cliff}).
\cref{subsec:gp_free_sets} treats $\GenericSet\subseteq\GeneralPermutation$ by applying~\cref{lemma:template} directly to constructions whose distance to all of $\GeneralPermutation$ is a closed form in $\FRACSAT{\varphi}$ (\cref{theorem:hardness_approx_non_superpos_sp_op,theorem:approx_non_superpos_hardness}).
\cref{subsec:assembly} combines these with~\cref{theorem:approx_identity_promise_hardness} to prove~\cref{theorem:approx_member_hardness} and~\cref{corollary:approx_opt_hardness}.

\subsection{Subsets of the Clifford-Implementable Unitaries} \label{subsec:id_to_general_free_sets}

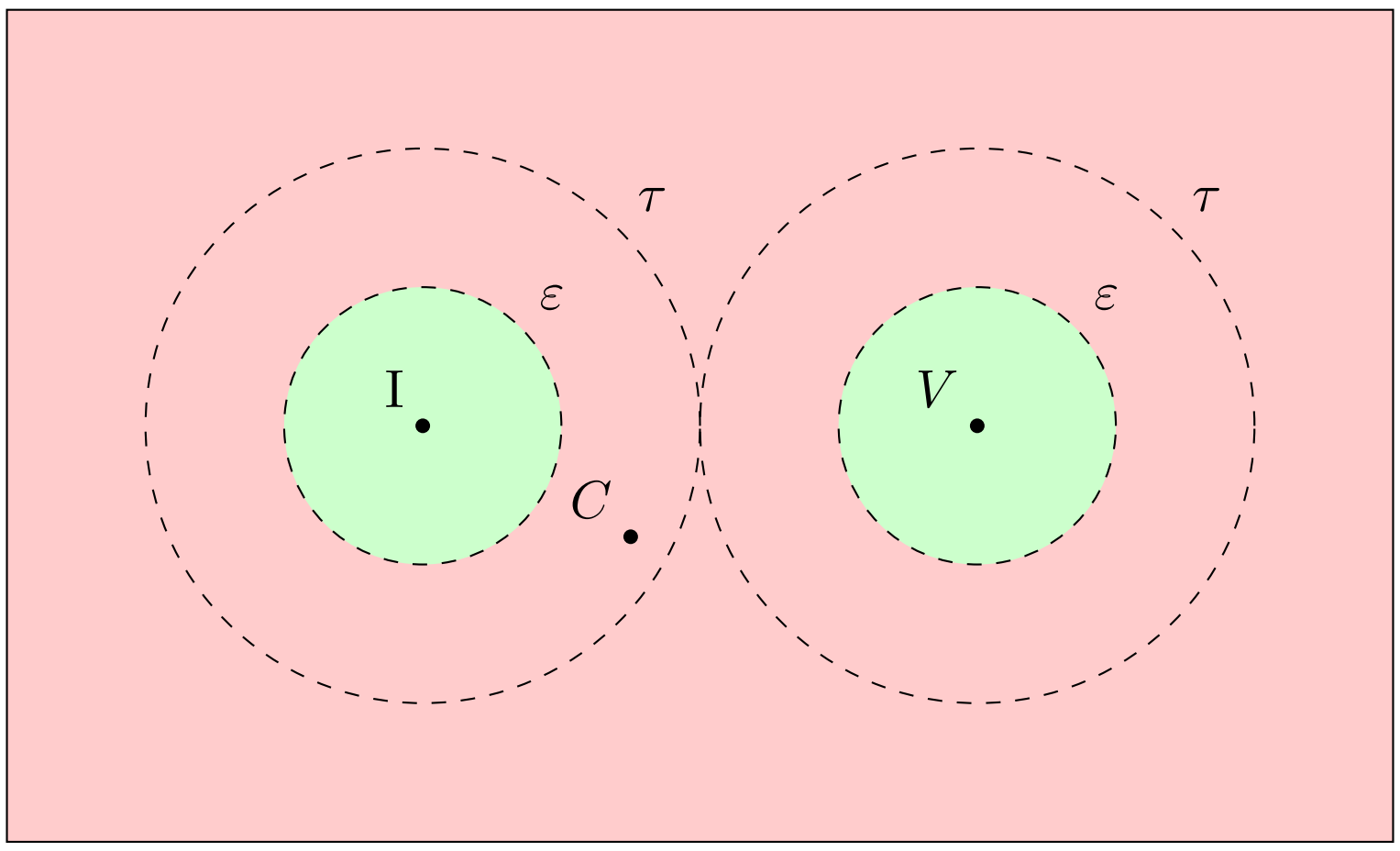
\begin{figure}
    \centering

    \begin{tikzpicture}
      \coordinate (I) at (0,0);
      \coordinate (C) at (4,0);
      \coordinate (C2) at (1.5,-0.8);
    
      \fill[red!20] (-3,-3) rectangle (7,3);       
      
      \fill[green!20] (I) circle (1);                     
      \fill[green!20] (C) circle (1);                     
    
      \draw (-3,-3) rectangle (7,3);
    
      \draw[dashed] (I) circle (1);
      \node at (45:1) [above right] {$\varepsilon$};
        
      \draw[dashed] (C) circle (1);
      \node at ($(C) + (45:1)$) [above right] {$\varepsilon$};

      \fill (I) circle (1.5pt) node[above left] {$\IGate$};
      \fill (C) circle (1.5pt) node[above left] {$V$};
      \fill (C2) circle (1.5pt) node[above left] {$C$};

      \draw[dashed] (I) circle (2);
      \node at (45:2) [above right] {$\tau$};

      \draw[dashed] (C) circle (2);
      \node at ($(C) + (45:2)$) [above right] {$\tau$};

    \end{tikzpicture}
    
    \caption{
    The triangle inequality argument of~\cref{lemma:reduction_promise_approx_id_to_approx_member}. Here $V\in \GenericSet$ lies at non-zero distance from $\IGate\in\GenericSet$, so $\Metric{\IGate}{V}{m} \geq \FullDist{\GenericSet}{m} \geq 2\tau$. The green regions contain the yes-instances of $\ApproxMember(\GenericSet,m,\varepsilon,\gateset)$ and the red region its no-instances; the dashed circle of radius $\tau$ around $\IGate$ bounds the promise.
    A no-instance $C$ of $\PromiseApproxIdentity(m,\varepsilon,\tau,\gateset)$ satisfies $\varepsilon < \Metric{C}{\IGate}{m} \leq \tau$ (by the promise). The triangle inequality then gives $\Metric{C}{V}{m} \geq \Metric{\IGate}{V}{m}- \Metric{C}{\IGate}{m} \geq 2\tau - \tau = \tau > \varepsilon$, so $\Metric{C}{\GenericSet}{m} > \varepsilon$.
    }
    \label{fig:approx_member_visualization}
\end{figure}

We prove that $\ApproxMember(\GenericSet,m,\varepsilon,\gateset)$ is $\PP$-hard when $\IGate\in\GenericSet\subseteq \SCliff$ by determining a range of tolerances $\varepsilon$ for which the problem coincides with $\PromiseApproxIdentity$, thereby inheriting the hardness of the latter.
To formalize this, let
$$\FullDist{\GenericSet}{m} := \inf\{ \Metric{\IGate}{U}{m} \colon U \in \GenericSet \wedge \Metric{\IGate}{U}{m} > 0 \}$$
denote the smallest non-zero distance with respect to $m$ between the identity and an element of $\GenericSet$, with $\inf(\emptyset):=\infty$.
Note that, for any two sets of unitaries $\GenericSet$ and $\GenericSet'$, if $\GenericSet\subseteq\GenericSet'$ then $\FullDist{\GenericSet}{m}\geq\FullDist{\GenericSet'}{m}$.
\cref{lemma:reduction_promise_approx_id_to_approx_member} shows that it suffices to establish lower-bounds on $\FullDist{\GenericSet}{m}$ in order to obtain hardness of $\ApproxMember(\GenericSet,m,\varepsilon,\gateset)$ (see \cref{fig:approx_member_visualization} for a pictorial illustration).

\smallskip

\begin{lemma}
    \label{lemma:reduction_promise_approx_id_to_approx_member}
    Consider any gate-set $\gateset$.
    Let $\GenericSet$ be a set of unitaries s.t.\ $\IGate \in \GenericSet$, let $m$ be a metric, and let $0<\varepsilon<\tau\leq\FullDist{\GenericSet}{m}/2$.
    Then every circuit $C$ over $\gateset$ with $\Metric{\IGate}{C}{m}\leq\tau$ is a yes-instance of $\PromiseApproxIdentity(m,\varepsilon,\tau,\gateset)$ iff $C$ is a yes-instance of $\ApproxMember(\GenericSet,m,\varepsilon,\gateset)$.
\end{lemma}

\begin{proof}
    Let $C$ be a circuit over $\gateset$ with $\Metric{\IGate}{C}{m}\leq\tau$.

    If $C$ is a yes-instance of $\PromiseApproxIdentity(m,\varepsilon,\tau,\gateset)$, then $\Metric{C}{\IGate}{m} \leq \varepsilon$, and since $\IGate\in \GenericSet$, the element $V:=\IGate$ of $\GenericSet$ satisfies $\Metric{C}{V}{m} \leq \varepsilon$.
    Hence $C$ is a yes-instance of $\ApproxMember(\GenericSet,m,\varepsilon,\gateset)$.

    If $C$ is a no-instance of $\PromiseApproxIdentity(m,\varepsilon,\tau,\gateset)$, then $\Metric{C}{\IGate}{m} > \varepsilon$, while $\Metric{C}{\IGate}{m} \leq \tau \leq \FullDist{\GenericSet}{m}/2$.
    We show that for all $V\in \GenericSet$, we have $\Metric{C}{V}{m} > \varepsilon$.
    If $\Metric{\IGate}{V}{m} = 0$, the triangle inequality gives $\Metric{C}{V}{m} \geq \Metric{C}{\IGate}{m} - \Metric{\IGate}{V}{m} = \Metric{C}{\IGate}{m} > \varepsilon$.
    Otherwise $\Metric{\IGate}{V}{m} \geq \FullDist{\GenericSet}{m}$, by definition of $\FullDist{.}{.}$.
    Again by the triangle inequality,
    \begin{align*}
        \Metric{C}{V}{m} 
            &\geq \Metric{\IGate}{V}{m} - \Metric{\IGate}{C}{m} \\
            &\geq \FullDist{\GenericSet}{m} - \Metric{\IGate}{C}{m} \\
            &\geq \FullDist{\GenericSet}{m} - \FullDist{\GenericSet}{m}/2 \\
            &= \FullDist{\GenericSet}{m}/2 \geq \tau > \varepsilon \ .
    \end{align*}
    Hence for all $V\in \GenericSet$, we have $\Metric{C}{V}{m} > \varepsilon$, meaning $C$ is a no-instance of $\ApproxMember(\GenericSet,m,\varepsilon,\gateset)$, as desired.
\end{proof}

In light of~\cref{lemma:reduction_promise_approx_id_to_approx_member}, we establish lower-bounds on $\FullDist{\GenericSet}{m}$ with $\GenericSet=\SCliff$ and $m$ being the Sine, spectral, or operator metric.
These lower-bounds allow us to select appropriate $\varepsilon$ for which $\ApproxMember(\GenericSet,m,\varepsilon,\gateset)$ is $\PP$-hard for every $\GenericSet$ with $\IGate\in\GenericSet\subseteq\SCliff$.
The following lemma lower-bounds $\FullDist{\SCliff}{m}$ for the three metrics $m$ we consider.

\smallskip
\begin{lemma}
    \label{lemma:full_dist_clifford}
    We have
    $$\FullDist{\SCliff}{\SineMetric}, \FullDist{\SCliff}{\SpectralAbbr}, \FullDist{\SCliff}{\OperatorAbbr} \geq \sqrt{2}/2 \ .$$
\end{lemma}

\begin{proof}
    We prove the bound for each metric.

    \ProofParagraph{Sine metric.}
    We prove that $\FullDist{\SCliff}{\SineMetric} \geq \sqrt{2}/2$.
    Minimizing $\Metric{\IGate}{U}{\SineMetric}=\sqrt{1-|\Tr(U)/2^n|^2}>0$ for an $n$-qubit Clifford unitary $U$ is equivalent to maximizing $|\Tr(U)/2^n|^2<1$.
    Let $\ket{\Phi} := 2^{-n/2}\sum_{x}\ket{x}\otimes\ket{x}$, with $x,y$ ranging over basis states in $\{0,1\}^n$ here and below. Then
    \begin{align*}
        |\Tr(U)/2^n|^2
            &= \Big|\frac{1}{2^n}\sum_{x}\bra{x}U\ket{x}\Big|^2 \\
            &= \Big|\frac{1}{2^n}\sum_{x,y}\braket{x|y}\bra{x}U\ket{y}\Big|^2
                & \left[\braket{x|y}=\delta_{xy}\right] \\
            &= \Big|\frac{1}{2^n}\sum_{x,y}(\bra{x}\otimes\bra{x})(\ket{y}\otimes U\ket{y})\Big|^2
                & \left[(\bra{a}\otimes\bra{b})(\ket{c}\otimes\ket{d})=\braket{a|c}\braket{b|d}\right] \\
            &= |\bra{\Phi}(\IGate\otimes U)\ket{\Phi}|^2 \ .
    \end{align*}

    This is the squared overlap of the $2n$-qubit stabilizer states $\ket{\Phi}$ and $(\IGate\otimes U)\ket{\Phi}$, where the latter is a stabilizer state since $\IGate\otimes U$ is Clifford.
    By Theorem 1 of~\cite{wagner2025unitary}, which builds on~\cite{garcia2014geometry}, for $n$-qubit stabilizer states $\ket{\psi_1}$ and $\ket{\psi_2}$, we have that either $|\braket{\psi_1|\psi_2}|^2=0$ or $|\braket{\psi_1|\psi_2}|^2=2^{-k}$ for $k\in \{0,1,\dots,n\}$.
    As our states are over $2n$ qubits, either $|\Tr(U)/2^n|^2 = 0$ or $|\Tr(U)/2^n|^2=2^{-k}$ for $k\in \{0,1,\dots,2n\}$.
    Maximizing $|\Tr(U)/2^n|^2<1$ yields $|\Tr(U)/2^n|^2=2^{-1}=1/2$.
    Plugging this in, we obtain 
    $\FullDist{\SCliff}{\SineMetric} \geq \sqrt{1-|\Tr(U)/2^n|^2} = \sqrt{1-1/2} = \sqrt{2}/2$.

    \ProofParagraph{Operator metric.}
    We prove that $\FullDist{\SCliff}{\OperatorAbbr} \geq \sqrt{2}/2$.
    Equivalently, we show that there is no Clifford unitary $U$ s.t.\ $0<\Metric{\IGate}{U}{\OperatorAbbr}<\sqrt{2}/2$.

    Suppose for contradiction that such a $U$ exists.
    Let $\alpha$ be s.t.\ $\OperatorNorm{U-e^{i\alpha}\IGate} = \Metric{U}{\IGate}{\OperatorAbbr}$, and let $P$ be a Pauli.
    Then
    \begin{align*}
        \OperatorNorm{U P U^\dagger -P}
            &= \OperatorNorm{UPU^\dagger-e^{i\alpha}PU^\dagger + e^{i\alpha}PU^\dagger-P} \\
            &\leq \OperatorNorm{UPU^\dagger-e^{i\alpha}PU^\dagger} + \OperatorNorm{e^{i\alpha}PU^\dagger-P} \\
            &= \OperatorNorm{U-e^{i\alpha}\IGate} + \OperatorNorm{e^{i\alpha} U^\dagger - \IGate} 
            & \left[\text{multiply by $UP^\dagger$ and $P^\dagger$}\right]\\
            &= \OperatorNorm{U-e^{i\alpha}\IGate} + \OperatorNorm{e^{-i\alpha} U - \IGate} & \left[\OperatorNorm{A^\dagger}=\OperatorNorm{A}\right] \\
            &= \OperatorNorm{U-e^{i\alpha}\IGate} + \OperatorNorm{U - e^{i\alpha}\IGate} \\
            &= 2\Metric{U}{\IGate}{\OperatorAbbr} \\
            &< 2(\sqrt{2}/2)=\sqrt{2} \ .
    \end{align*}
    By the definition of the Clifford group, we see that $U P U^\dagger$ is a Pauli.
    For any Paulis $P$ and $Q$, we have $\OperatorNorm{Q -P} \in \{0,\sqrt{2},2\}$.
    From the above, we have $\OperatorNorm{U P U^\dagger -P} < \sqrt{2}$, so in particular, $\OperatorNorm{U P U^\dagger -P}=0$, meaning $U P U^\dagger = P$. Since we made no assumptions on $P$, this holds for all Paulis $P$.
    For $U$ satisfying $U P U^\dagger = P$ for all $P$, we have $U=e^{i\beta}\IGate$ for some $\beta$, but then $\Metric{\IGate}{U}{\OperatorAbbr}=0$, yielding a contradiction.

    \ProofParagraph{Spectral metric.}
    We prove that $\FullDist{\SCliff}{\SpectralAbbr} \geq \sqrt{2}/2$. 
    Let $U \in \SCliff$ with $\Metric{\IGate}{U}{\SpectralAbbr} > 0$. 
    If $\Metric{\IGate}{U}{\OperatorAbbr} > 0$ then $\Metric{\IGate}{U}{\SpectralAbbr} \geq \Metric{\IGate}{U}{\OperatorAbbr} \geq \FullDist{\SCliff}{\OperatorAbbr} \geq \sqrt{2}/2$. 
    Otherwise $\Metric{\IGate}{U}{\OperatorAbbr} = 0$, that is, $U = e^{i\alpha}\IGate$, and $\alpha \neq 0$ since $\Metric{\IGate}{U}{\SpectralAbbr} > 0$. 
    Every entry of $\HGate$, $\SGate$, and $\CXGate$ lies in the ring $\Zring{1/\sqrt{2}, i}$.
    Hence every entry of the matrix of a circuit $C$ over these gates, and thus of $U_C$, a block of that matrix, lies in this ring. 
    As $U=U_C$ for some such $C$, we have $e^{i\alpha} \in \Zring{1/\sqrt{2}, i}$. 
    The elements of modulus 1 in this ring are the 8th roots of unity~\cite[Appendix~A]{kliuchnikov2013fast}, so $\alpha$ is a non-zero multiple of $\pi/4$ and $\Metric{\IGate}{e^{i\alpha}\IGate}{\SpectralAbbr} = 2|\sin(\alpha/2)| \geq 2\sin(\pi/8) > \sqrt{2}/2$.
\end{proof}

Combining~\cref{lemma:reduction_promise_approx_id_to_approx_member,lemma:full_dist_clifford}, we obtain hardness for every $\GenericSet$ with $\IGate\in\GenericSet\subseteq\SCliff$.

\smallskip
\begin{theorem}
    \label{theorem:hardness_approx_cliff}
    Consider any gate-set $\gateset$ that exactly implements Clifford+$\TGate$. 
    Let $m$ be the Sine, spectral, or operator metric, and let $\GenericSet$ be a set of unitaries with $\IGate\in\GenericSet\subseteq\SCliff$. 
    Fix $\PTIME$-time computable $0<\varepsilon<\sqrt{2}/4$.
    Then $\ApproxMember(\GenericSet,m,\varepsilon,\gateset)$ is $\PP$-hard.
\end{theorem}

\begin{proof}
    Let $0<\varepsilon<\sqrt{2}/4$ and set $\tau:=\sqrt{2}/4$.
    As $\GenericSet\subseteq\SCliff$, \cref{lemma:full_dist_clifford} gives $\FullDist{\GenericSet}{m}\geq\FullDist{\SCliff}{m}\geq \sqrt{2}/2 = 2\tau$.
    Each $m$ is a metric and $\IGate\in\GenericSet$, so by~\cref{lemma:reduction_promise_approx_id_to_approx_member}, every circuit $C$ over $\gateset$ with $\Metric{\IGate}{C}{m}\leq\tau$ is a yes-instance of $\PromiseApproxIdentity(m,\varepsilon,\tau,\gateset)$ iff $C$ is a yes-instance of $\ApproxMember(\GenericSet,m,\varepsilon,\gateset)$.
    Since $\tau < 1 \leq D_m$, \cref{theorem:approx_identity_promise_hardness} shows that $\PromiseApproxIdentity(m,\varepsilon,\tau,\gateset)$ is $\PP$-hard, and hence so is $\ApproxMember(\GenericSet,m,\varepsilon,\gateset)$.
\end{proof}

\subsection{Subsets of the Generalized Permutation Matrices} \label{subsec:gp_free_sets}

We now consider sets of unitaries $\GenericSet$ with $\IGate\in\GenericSet\subseteq\GeneralPermutation$.
The strategy of~\cref{subsec:id_to_general_free_sets} does not apply here.
For example, consider the unitary $\diag(1,\dots,1,e^{i\theta})\in\GeneralPermutation$, parameterized by $\theta\in\reals$.
Its distance to $\IGate$ tends to $0$ as $\theta\to 0$ under each of the three metrics, hence $\FullDist{\GeneralPermutation}{m}=0$, in contrast to \cref{lemma:full_dist_clifford} for $\SCliff$.
Instead, we construct circuits whose distance to $\GenericSet$ is a closed form in $\FRACSAT{\varphi}$ and apply~\cref{lemma:template} directly.
For the Sine metric and a constant $\ell$, $C$ applies $(\HGate\ZGate)^{\otimes\ell}$ controlled on $\varphi$, and $\Metric{\GenericSet}{C}{\SineMetric}$ follows from a bound on $|\Tr(U^\dagger C)|$ over $U\in\GeneralPermutation$.
For the spectral and operator metrics, $\ApproxMember(\GenericSet,m,\varepsilon,\gateset)$ and $\ApproxIdentity(m,\varepsilon,\gateset)$ still agree on the circuits $C$ of~\cref{lemma:spectral_construction} for which $\FRACSAT{\varphi}$ is sufficiently small, as $\Metric{\GenericSet}{C}{m}=\Metric{\IGate}{C}{m}$ for these circuits.

\Paragraph{Sine metric.}
The following lemma expresses the distance of a $\varphi$-controlled $(\HGate\ZGate)^{\otimes \ell}$ circuit from any generalized permutation matrix as a function of $\FRACSAT{\varphi}$.

\smallskip
\begin{lemma}
    \label{lemma:superpos_construction}
    Let $\varphi$ be a propositional formula over $n$ bits, let $\ell\geq 1$, let $\GenericSet$ satisfy $\IGate \in \GenericSet \subseteq \GeneralPermutation$, and let $C:= C_\varphi((\HGate\ZGate)^{\otimes \ell})$ be the $(n+\ell)$-qubit unitary with $C \colon \ket{x,y} \mapsto \ket{x}((\HGate\ZGate)^{\otimes \ell}\ket{y})$ when $\varphi(x)$ and $C\colon \ket{x,y} \mapsto \ket{x,y}$ otherwise.
    Then
    $$\Metric{\GenericSet}{C}{\SineMetric} = \sqrt{1-(1-(1-2^{-\ell/2})\FRACSAT{\varphi})^2} \ .$$
\end{lemma}

\begin{proof}
    We derive an expression for
    \begin{align*}
        \Metric{\GenericSet}{C}{\SineMetric} 
            &= \sqrt{1- \big(\sup_{U\in\GenericSet} |\Tr(U^\dagger C)|/2^{n+\ell}\big)^2} \ .
    \end{align*}
    We do so by giving an upper bound on $\sup_{U\in\GenericSet} |\Tr(U^\dagger C)|$ and showing a specific $U$ that attains this bound.
    Since $\GenericSet \subseteq \GeneralPermutation$ and $\GeneralPermutation$ is closed under taking adjoints, $\sup_{U \in \GenericSet}|\Tr(U^\dagger C)| \leq \sup_{U \in \GeneralPermutation}|\Tr(UC)|$. 
    We bound the latter and show that $\IGate \in \GenericSet$ attains it.

    Towards the upper bound, 
    let $U \in \GeneralPermutation$, so that $U \colon \ket{x,y} \mapsto e^{ih(x,y)} \ket{L(x,y)}$ for a bijection $L$ on $\{0,1\}^{n+\ell}$ and a phase function $h$.
    For $x$ s.t.\ $\neg\varphi(x)$, we have $\braket{x,y|U C|x,y}=\braket{x,y|U|x,y}=e^{ih(x,y)}\braket{x,y|L(x,y)}$, so $|\braket{x,y|U C|x,y}| \leq 1$. 
    For $x$ s.t.\ $\varphi(x)$, we have 
    $$\braket{x,y|UC|x,y}=\sum_z c_{zy} \braket{x,y|U|x,z} = \sum_z c_{zy} e^{ih(x,z)}\braket{x,y|L(x,z)} \ ,$$ 
    where $c_{zy}:=\braket{z|(\HGate\ZGate)^{\otimes \ell}|y}$ satisfies $|c_{zy}|=2^{-\ell/2}$, since every entry of $\HGate\ZGate$ has modulus $2^{-1/2}$.
    Since $L$ is a bijection, at most one summand is non-zero, so $|\braket{x,y|U C|x,y}|\leq 2^{-\ell/2}$.
    Then
    \begin{align*}
        |\Tr(UC)| 
            &= \Big|\sum_{x,y} \braket{x,y|U C|x,y}\Big| \\
            &\leq \sum_{x,y} |\braket{x,y|U C|x,y}| \\
            &= \sum_{x,y \colon \neg \varphi(x)} |\braket{x,y|U C|x,y}|+\sum_{x,y \colon \varphi(x)} |\braket{x,y|U C|x,y}| \\
            &\leq 2^{n+\ell}\FRACSAT{\neg\varphi}+2^{n+\ell}\FRACSAT{\varphi}2^{-\ell/2} \ . 
    \end{align*}
    For a specific $U$ witnessing this bound, take $U := \IGate\in \GenericSet$.
    Since the diagonal entries of $\HGate\ZGate$ are both $2^{-1/2}$ and diagonals multiply under tensor products, $\braket{y|(\HGate\ZGate)^{\otimes \ell}|y}=2^{-\ell/2}$ for every $y$, so
    $$\Tr(UC) = \sum_{x,y \colon\neg\varphi(x)} 1 + \sum_{x,y \colon \varphi(x)} 2^{-\ell/2} = 2^{n+\ell} \FRACSAT{\neg\varphi} + 2^{n+\ell}\FRACSAT{\varphi}2^{-\ell/2} \ .$$
    The bound is therefore attained, and
    $$\Metric{\GenericSet}{C}{\SineMetric} = \sqrt{1-(\FRACSAT{\neg\varphi}+2^{-\ell/2}\FRACSAT{\varphi})^2} = \sqrt{1-(1-(1-2^{-\ell/2})\FRACSAT{\varphi})^2} \ .$$
\end{proof}

\smallskip
\begin{theorem}
    \label{theorem:approx_non_superpos_hardness}
    Consider any gate-set $\gateset$ that exactly implements Clifford+$\TGate$, and let $\GenericSet$ be a set of unitaries with $\IGate\in\GenericSet\subseteq\GeneralPermutation$. 
    Fix $\PTIME$-time computable $0<\varepsilon<1$. 
    Then $\ApproxMember(\GenericSet,\SineMetric,\varepsilon,\gateset)$ is $\PP$-hard.
\end{theorem}

\begin{proof}
    Let $0 < \varepsilon < 1$ be given. 
    Since $\sqrt{1 - 2^{-\ell}} \to 1$ as $\ell \to \infty$, fix $\ell \geq 1$ with $\tau := \sqrt{1 - 2^{-\ell}}> \varepsilon$. 
    Given $\varphi$ over $n$ bits, construct $C := C_\varphi ((\HGate\ZGate)^{\otimes\ell})$ over $n + \ell$ qubits as follows: apply $C_\varphi \XGate$\footnote{We implement $C_\varphi\XGate\colon \ket{x}\ket{b} \mapsto \ket{x}\ket{b \oplus \varphi(x)}$ by the ancilla-restoring Clifford+$\TGate$ circuit constructed for the Sine metric in~\cref{sec:lower_bounds_on_approx_ID}.} to write $\varphi(x)$ into a clean ancilla qubit; apply a $\ZGate$ and then an $\HGate$ gate to each of the $\ell$ target qubits, each controlled on the ancilla; and uncompute the ancilla with a second $C_\varphi \XGate$.
    All ancillas are restored, and a controlled-$\HGate$ gate is exactly implementable over Clifford+$\TGate$, so $C$ is a polynomially sized Clifford+$\TGate$ circuit and the mapping is $\PTIME$-time.

    Set $g(x) := \sqrt{1 - (1 - (1 - 2^{-\ell/2})x)^2}$ and $\kappa := 1$. 
    Then $g(0) = 0$, $g(\kappa) = \sqrt{1-(2^{-\ell/2})^2} = \sqrt{1 - 2^{-\ell}} = \tau > 0$, and $g$ is strictly increasing and continuous on $[0,1]$, since $1 - (1 - 2^{-\ell/2})x$ strictly decreases from $1$ to $2^{-\ell/2} > 0$ on $[0,1]$. 
    Since $\IGate\in \GenericSet\subseteq \GeneralPermutation$, \cref{lemma:superpos_construction} applies, so $g(\FRACSAT{\varphi}) = \Metric{\GenericSet}{C}{\SineMetric}$ for every $\varphi$.
    
    As $g^{-1}(\varepsilon) = (1-\sqrt{1-\varepsilon^2})/(1-2^{-\ell/2})$ is $\PTIME$-time computable, \cref{lemma:template} yields $\PP$-hardness of $\PromiseApproxMember(\GenericSet, \SineMetric, \varepsilon, \tau, \gateset)$, and since $\PromiseApproxMember(\GenericSet, \SineMetric, \varepsilon, \tau, \gateset)$ and $\ApproxMember(\GenericSet, \SineMetric, \varepsilon, \gateset)$ agree on all circuits satisfying the promise of the former, we have $\PP$-hardness of $\ApproxMember(\GenericSet, \SineMetric, \varepsilon, \gateset)$. 
    As $\ell$ was chosen with $\tau > \varepsilon$ for any $\varepsilon < 1$, every $\PTIME$-time computable $0 < \varepsilon < 1$ is covered. 
\end{proof}

\Paragraph{Spectral and operator metrics.}
The following lemma shows that, for the circuit $C$ of~\cref{lemma:spectral_construction}, the closest element of $\GeneralPermutation$ is $\IGate$ when $\FRACSAT{\varphi}$ is sufficiently small.

\smallskip
\begin{lemma}
    \label{lemma:spectral_GP_distance}
    Let $\varphi$ be a propositional formula over $n \geq 5$ bits, and let $C$ be the circuit constructed from $\varphi$ in~\cref{lemma:spectral_construction}.
    If $\FRACSAT{\varphi} \leq \frac{1}{2}\left(1 - \sqrt{\frac{1 + 2^{4-n}}{2}}\right)$, then
    $$\Metric{C}{\GeneralPermutation}{\SpectralAbbr} = \Metric{C}{\IGate}{\SpectralAbbr} \ .$$
\end{lemma}

\begin{proof}
    As $\IGate\in \GeneralPermutation$, we have $\Metric{C}{\IGate}{\SpectralAbbr} \geq \Metric{C}{\GeneralPermutation}{\SpectralAbbr}$, so it suffices to prove $\Metric{C}{\IGate}{\SpectralAbbr} \leq \Metric{C}{\GeneralPermutation}{\SpectralAbbr}$.

    Recall that $\theta=2 \sin^{-1}\sqrt{\FRACSAT{\varphi}}$, so $\FRACSAT{\varphi} = (1-\cos\theta)/2$ and the assumption is equivalent to $\cos\theta \geq \sqrt{(1+2^{4-n})/2}$.
    As the right-hand side is at least $1/\sqrt{2}$ and $\theta\in[0,\pi]$, this gives $\theta \in [0, \pi/4]$, so $\sin\theta \geq 0$ and $\sin 2\theta \geq 0$, and squaring gives $\cos 2\theta = 2\cos^2\theta - 1 \geq 2^{4-n} > 0$.
    If $\theta = 0$, then $U_C=\IGate$ by~\cref{lemma:spectral_construction}, so both sides are $0$.
    Hence we assume that $0 < \theta \leq \pi/4$, so that both $\varphi$ and $\neg\varphi$ are satisfiable and $\ket{\varphi}$ and $\ket{\neg\varphi}$ are well-defined.
    By~\cref{lemma:spectral_construction},
    $$U_C = \IGate - 2\sin^2(\theta)\big(\ket{\neg\varphi}\bra{\neg\varphi} + \ket{\varphi}\bra{\varphi}\big)
            + \sin (2\theta)\big(\ket{\varphi}\bra{\neg\varphi} - \ket{\neg\varphi}\bra{\varphi}\big) \ .$$

    We expand $\Metric{C}{\IGate}{\SpectralAbbr}$ and $\Metric{C}{\GeneralPermutation}{\SpectralAbbr}$.
    By~\cref{lemma:spectral_construction}, the eigenangles of $U_C$ are $\{0,\pm 2\theta\}$, so from the definition of the spectral metric, we see that $\Metric{C}{\IGate}{\SpectralAbbr} = 2\max\{|\sin\pm \theta|,|\sin 0|\} = 2 \sin \theta$.
    Let $\ket{\psi}$ range over unit vectors.
    We will use the general fact that $\OperatorNorm{U-V}=\sup_{\ket\psi}\sqrt{2 - 2\Re(\braket{\psi|U^\dagger V|\psi})}$. Then we have
    \begin{align*}
        \Metric{C}{\GeneralPermutation}{\SpectralAbbr} &= \inf_{U\in \GeneralPermutation} \sup_{\ket{\psi}} ||(U_C-U)\ket{\psi}||_2 \\
        &= \inf_{U\in \GeneralPermutation} \sup_{\ket{\psi}} \sqrt{2-2\Re(\braket{\psi|U^\dagger U_C|\psi})} \ .
    \end{align*}
    Hence we need to prove
    $$2 \sin \theta \leq \inf_{U\in \GeneralPermutation} \sup_{\ket{\psi}} \sqrt{2-2\Re(\braket{\psi|U^\dagger U_C|\psi})} \ .$$
    It suffices to show that for any $U\in \GeneralPermutation$, there is a choice of $\ket{\psi}$ s.t.\ $2 \sin \theta \leq \sqrt{2-2\Re(\braket{\psi|U^\dagger U_C|\psi})}$, that is, $\Re(\braket{\psi|U^\dagger U_C|\psi}) \leq 1 - 2\sin^2\theta = \cos 2\theta$.
    We do so by case distinction on $U$, where $U\ket{x} = e^{ih(x)}\ket{L(x)}$.

    If $L$ is not the identity, then pick $\ket{\psi}:=\ket{x_0}$ where $x_0$ is s.t.\ $L(x_0)\neq x_0$.
    Then
    \begin{align*}
        \Re(\braket{x_0|U^\dagger U_C|x_0}) = \Re(e^{-ih(x_0)}\braket{L(x_0)|U_C|x_0}) \leq |\braket{L(x_0)|U_C|x_0}| \ .
    \end{align*}
    We upper-bound $|\braket{y|U_C|x}|$ for all $x \neq y$.
    As $\braket{y|\IGate|x} = 0$, the triangle inequality gives
    \begin{align*}
        |\braket{y|U_C|x}| &\leq 
      2\sin^2(\theta)\big(|\braket{y|\neg\varphi}\braket{\neg\varphi|x}| + |\braket{y|\varphi}\braket{\varphi|x}|\big) \\
      & \qquad + \sin(2\theta)\big(|\braket{y|\varphi}\braket{\neg\varphi|x}| + |\braket{y|\neg\varphi}\braket{\varphi|x}|\big) \ .
    \end{align*}
    For all $z$ and $\varphi$, we have $|\braket{z|\varphi}| \leq \tfrac{1}{\sqrt{2^n\FRACSAT{\varphi}}}$. 
    Since $\sin^2(\theta)\geq 0$ and $\sin(2\theta)\geq0$, we have
    \begin{align*}
        |\braket{y|U_C|x}| \leq 2\sin^2(\theta) \Big(\frac{1}{2^n\FRACSAT{\neg\varphi}} + \frac{1}{2^n\FRACSAT{\varphi}}\Big)  + \sin(2\theta) \frac{2}{2^n \sqrt{\FRACSAT{\varphi}\FRACSAT{\neg\varphi}}} \ .
    \end{align*}
    As $\FRACSAT{\varphi} = \sin^2(\theta/2)$ and $\FRACSAT{\neg\varphi} = \cos^2(\theta/2)$, we have $\sin\theta = 2\sin(\theta/2)\cos(\theta/2) = 2\sqrt{\FRACSAT{\varphi}\FRACSAT{\neg\varphi}}$, so $\sin^2\theta = 4\FRACSAT{\varphi}\FRACSAT{\neg\varphi}$ and $\sin 2\theta = 2\sin\theta\cos\theta = 4\sqrt{\FRACSAT{\varphi}\FRACSAT{\neg\varphi}}\cos\theta$.
    Substituting, we obtain
    \begin{align*}
        |\braket{y|U_C|x}|
        &\leq \frac{8\FRACSAT{\varphi}\FRACSAT{\neg\varphi}}{2^n}\Big(\frac{1}{\FRACSAT{\neg\varphi}} + \frac{1}{\FRACSAT{\varphi}}\Big) + \frac{8\cos\theta}{2^n} \\
        &= 2^{3-n}\big(\FRACSAT{\varphi} + \FRACSAT{\neg\varphi}\big) + 2^{3-n}\cos\theta \\
        &\leq 2^{4-n} \ ,
    \end{align*}
    where the final step uses $\FRACSAT{\varphi} + \FRACSAT{\neg\varphi} = 1$ and $\cos\theta \leq 1$.
    Hence $\Re(\braket{x_0|U^\dagger U_C|x_0}) \leq 2^{4-n} \leq \cos 2\theta$.

    If $L$ is the identity, we pick $\ket{\psi}$ as one of the two eigenstates
    $\ket{\psi_\pm} := \frac{1}{\sqrt{2}}(\ket{\varphi} \pm i\ket{\neg\varphi})$ of $U_C$, which satisfy $U_C\ket{\psi_\pm} = e^{\pm 2i\theta}\ket{\psi_\pm}$.
    Note that, as $U$ is diagonal, $U^\dagger\ket{\varphi}$ has support exactly on $\{x\colon \varphi(x)\}$, so in particular, $\braket{\varphi|U^\dagger|\neg\varphi} = \braket{\neg\varphi|U^\dagger|\varphi} = 0$.
    Let $\beta := \frac{1}{2}\big(\braket{\varphi|U^\dagger|\varphi} + \braket{\neg\varphi|U^\dagger|\neg\varphi}\big)$, and note that $\Re(\beta) \leq |\beta| \leq 1$, as $|\braket{\varphi|U^\dagger|\varphi}|, |\braket{\neg\varphi|U^\dagger|\neg\varphi}| \leq 1$.
    Then
    \begin{align*}
        \Re(\braket{\psi_\pm|U^\dagger U_C|\psi_\pm})
        &= \Re\big(e^{\pm 2i\theta}\braket{\psi_\pm|U^\dagger|\psi_\pm}\big) \\
        &= \Re\big(\frac{e^{\pm 2i\theta}}{2}(\braket{\varphi|U^\dagger|\varphi} \pm i \braket{\varphi|U^\dagger|\neg\varphi} \mp i \braket{\neg\varphi|U^\dagger|\varphi} + \braket{\neg\varphi|U^\dagger|\neg\varphi})\big) \\
        &= \Re\big(\frac{e^{\pm 2i\theta}}{2}(\braket{\varphi|U^\dagger|\varphi} + \braket{\neg\varphi|U^\dagger|\neg\varphi})\big) \\
        &= \Re\big(e^{\pm 2i\theta}\beta\big) \\
        &= \cos(2\theta) \Re(\beta) \mp \sin(2\theta) \Im(\beta) \ .
    \end{align*}
    If $\Im(\beta) \geq 0$, then pick $\ket{\psi} := \ket{\psi_+}$, and otherwise pick $\ket{\psi} := \ket{\psi_-}$.
    This ensures that the second term equals $-\sin(2\theta)|\Im(\beta)|$, and since $\sin 2\theta \geq 0$, $\cos 2\theta \geq 0$, and $\Re(\beta) \leq 1$, we have
    $$\Re(\braket{\psi|U^\dagger U_C|\psi}) = \cos(2\theta) \Re(\beta) - \sin(2\theta)|\Im(\beta)| \leq \cos(2\theta) \Re(\beta) \leq \cos 2\theta \ .$$
    
    Hence in both cases, $\Re(\braket{\psi|U^\dagger U_C|\psi}) \leq \cos 2\theta$ for some choice of $\ket{\psi}$, as desired.
\end{proof}

Hardness of $\ApproxMember$ then follows from the fact that it coincides with $\ApproxIdentity$ on the circuits $C$ of~\cref{lemma:spectral_construction} for which $\varphi$ satisfies the condition of~\cref{lemma:spectral_GP_distance}.

\smallskip
\begin{theorem}
    \label{theorem:hardness_approx_non_superpos_sp_op}
    Consider any gate-set $\gateset$ that exactly implements Clifford+$\TGate$, and let $\GenericSet$ be a set of unitaries with $\IGate\in\GenericSet\subseteq\GeneralPermutation$.
    Fix $\PTIME$-time computable $0<\varepsilon<\sqrt{2}$.
    Then $\ApproxMember(\GenericSet, \SpectralAbbr, \varepsilon,\gateset)$ and $\ApproxMember(\GenericSet, \OperatorAbbr, \varepsilon, \gateset)$ are $\PP$-hard.
\end{theorem}

\begin{proof}
    Let $m$ be the spectral or the operator metric, and let $0 < \varepsilon < \sqrt{2}$ be given.

    Fix $\tau$ with $\varepsilon < \tau < \sqrt{2}$.
    Set $g(x) := 4\sqrt{x(1-x)}$ and $\kappa := g^{-1}(\tau) = (1 - \sqrt{1 - \tau^2/4})/2$.
    As $g(\sin^2(\pi/8)) = \sqrt{2} > \tau$, we have $\kappa < \sin^2(\pi/8)$.
    Then $g(0) = 0$, $g(\kappa) = \tau > 0$, and $g$ is strictly increasing and continuous on $[0, \kappa]$.
    Since $\frac{1}{2}\big(1 - \sqrt{(1 + 2^{4-\ell})/2}\big) \to \frac{1}{2}\big(1 - 1/\sqrt{2}\big) = \sin^2(\pi/8)$ as $\ell \to \infty$, fix $\ell \geq 5$ with
    $\kappa \leq \tfrac{1}{2}\left(1 - \sqrt{(1 + 2^{4-\ell})/2}\right)$.

    Given $\varphi$ over $n$ bits, let $\varphi'(x, y) := \varphi(x)$, where $x$ ranges over $n$ bits and $y$ over $\ell$ bits, so that $\FRACSAT{\varphi'} = \FRACSAT{\varphi}$.
    Construct $C$ as in~\cref{lemma:spectral_construction} from $\varphi'$ over $n + \ell \geq 5$ bits.
    Since $\ell$ is constant, $C$ is a polynomially sized Clifford+$\TGate$ circuit, and the mapping is $\PTIME$-time.
    To apply~\cref{lemma:template}, it suffices to show that $g(\FRACSAT{\varphi}) = \Metric{\GenericSet}{C}{m}$ for $0 \leq \FRACSAT{\varphi} \leq \kappa$.

    Towards this, note that the bound on $\FRACSAT{\varphi}$ in~\cref{lemma:spectral_GP_distance} is increasing in the number of bits, so for $0 \leq \FRACSAT{\varphi} \leq \kappa$, we have
    $$\FRACSAT{\varphi'} = \FRACSAT{\varphi} \leq \kappa \leq \frac{1}{2}\left(1 - \sqrt{(1 + 2^{4-\ell})/2}\right) \leq \frac{1}{2}\left(1 - \sqrt{(1 + 2^{4-(n+\ell)})/2}\right) \ .$$
    Hence~\cref{lemma:spectral_GP_distance} applies on $[0, \kappa]$, where it yields $\Metric{C}{\GeneralPermutation}{\SpectralAbbr} = \Metric{C}{\IGate}{\SpectralAbbr}$.
    By~\cref{lemma:spectral_construction}, the eigenangles of $C$ are $\{0,\pm 2\theta(\varphi')\}$, so $\Metric{\IGate}{C}{\SpectralAbbr} = 2|\sin \theta(\varphi')| = 2|\sin \theta(\varphi)| = 2\sin \theta(\varphi) = 4\sqrt{\FRACSAT{\varphi}(1 - \FRACSAT{\varphi})} = g(\FRACSAT{\varphi})$, using $\FRACSAT{\varphi'} = \FRACSAT{\varphi}$, the definition of $\theta(\varphi)$, and standard trigonometry.
    We now prove $g(\FRACSAT{\varphi}) = \Metric{\GenericSet}{C}{m}$ for each of the two metrics.

    \ProofParagraph{Spectral metric.}
    As $\IGate \in \GenericSet \subseteq \GeneralPermutation$, we have
    $$\Metric{\IGate}{C}{\SpectralAbbr} \geq \Metric{\GenericSet}{C}{\SpectralAbbr} \geq \Metric{\GeneralPermutation}{C}{\SpectralAbbr} = \Metric{\IGate}{C}{\SpectralAbbr} \ ,$$
    so $g(\FRACSAT{\varphi}) = \Metric{\GenericSet}{C}{\SpectralAbbr}$.
    
    \ProofParagraph{Operator metric.}
    For every $U \in \GeneralPermutation$ and $\alpha \in \reals$, we have $e^{i\alpha}U \in \GeneralPermutation$, so
    $$\Metric{C}{U}{\OperatorAbbr} = \min_\alpha \OperatorNorm{C - e^{i\alpha}U} \geq \Metric{C}{\GeneralPermutation}{\SpectralAbbr} = g(\FRACSAT{\varphi}) \ ,$$
    and as $\GenericSet \subseteq \GeneralPermutation$, we see that $\Metric{\GenericSet}{C}{\OperatorAbbr} \geq g(\FRACSAT{\varphi})$.
    Conversely, we have $\Metric{\IGate}{C}{\OperatorAbbr} \leq \Metric{\IGate}{C}{\SpectralAbbr}$, so as $\IGate \in \GenericSet$,
    $$\Metric{\GenericSet}{C}{\OperatorAbbr} \leq \Metric{\IGate}{C}{\OperatorAbbr} \leq \Metric{\IGate}{C}{\SpectralAbbr} = g(\FRACSAT{\varphi}) \ .$$
    Hence $g(\FRACSAT{\varphi}) = \Metric{\GenericSet}{C}{\OperatorAbbr}$.

    As $g^{-1}(\varepsilon) = (1-\sqrt{1-\varepsilon^2/4})/2$ is $\PTIME$-time computable, \cref{lemma:template} yields $\PP$-hardness of $\PromiseApproxMember(\GenericSet, m, \varepsilon, \tau, \gateset)$, and since $\PromiseApproxMember(\GenericSet, m, \varepsilon, \tau, \gateset)$ and $\ApproxMember(\GenericSet, m, \varepsilon, \gateset)$ agree on all circuits satisfying the promise of the former, we have $\PP$-hardness of $\ApproxMember(\GenericSet, m, \varepsilon, \gateset)$.
    As $\tau$ was chosen with $\varepsilon < \tau < \sqrt{2}$ for any $\varepsilon < \sqrt{2}$, every $\PTIME$-time computable $0 < \varepsilon < \sqrt{2}$ is covered.
\end{proof}

\subsection{Proof of the Main Results} \label{subsec:assembly}

With the individual results in hand, we are ready to assemble~\cref{theorem:approx_member_hardness} and~\cref{corollary:approx_opt_hardness}.

\begin{proof}[Proof of~\cref{theorem:approx_member_hardness}]
    Let $\GenericSet$, $m$, and a $\PTIME$-time computable $\varepsilon$ be as in some cell of~\cref{table:all_problems_hardness2}.
    There are three cases.

    \ProofParagraph{$\GenericSet = \SId$.}
    The cell requires $0 < \varepsilon < D_m$. 
    Pick $\tau$ with $\varepsilon < \tau < D_m$, which is possible since $\varepsilon < D_m$. 
    By~\cref{theorem:approx_identity_promise_hardness}, $\PromiseApproxIdentity(m, \varepsilon, \tau, \gateset)$ is $\PP$-hard. 
    As $\PromiseApproxIdentity(m, \varepsilon, \tau, \gateset)$ and $\ApproxIdentity(m, \varepsilon, \gateset) = \ApproxMember(\SId, m, \varepsilon, \gateset)$ agree on all circuits satisfying the promise of the former, $\PP$-hardness follows.

    \ProofParagraph{$\IGate\in\GenericSet\subseteq\SCliff$.}
    The cell requires $0 < \varepsilon < \sqrt{2}/4$, and $\ApproxMember(\GenericSet, m, \varepsilon, \gateset)$ is $\PP$-hard for every such $\varepsilon$ by~\cref{theorem:hardness_approx_cliff}.

    \ProofParagraph{$\IGate\in\GenericSet\subseteq\GeneralPermutation$.}
    For $m$ the Sine metric, the cell requires $0 < \varepsilon < 1$, and 
    $\ApproxMember(\GenericSet, \SineMetric, \varepsilon, \gateset)$ 
    is $\PP$-hard for every such $\varepsilon$ by~\cref{theorem:approx_non_superpos_hardness}.
    For $m$ the spectral or operator metric, the cell requires $0 < \varepsilon < \sqrt{2}$, and 
    $\ApproxMember(\GenericSet, m, \varepsilon, \gateset)$ 
    is $\PP$-hard for every such $\varepsilon$ by~\cref{theorem:hardness_approx_non_superpos_sp_op}.

    For the boldface cells, whose range is $(0, D_m)$, no larger $\varepsilon$ can be added: every $\varepsilon \geq D_m$ makes $\ApproxMember(\GenericSet, m, \varepsilon, \gateset)$ accept all instances.
\end{proof}

\begin{proof}[Proof of~\cref{corollary:approx_opt_hardness}]
    Let $f$ be a cost function with free set $\GenericSet_f$, and let $m$ and $\varepsilon$ be as in the cell of~\cref{table:all_problems_hardness2} for $\GenericSet = \GenericSet_f$.
    We give a Karp reduction from $\ApproxMember(\GenericSet_f, m, \varepsilon,\gateset)$ to $\ApproxOpt(f,m,\varepsilon,\gateset)$. 
    Then by~\cref{theorem:approx_member_hardness}, $\ApproxMember(\GenericSet_f, m, \varepsilon,\gateset)$ is $\PP$-hard, so the result follows.

    Given a yes/no instance $C$ of $\ApproxMember(\GenericSet_f, m, \varepsilon,\gateset)$, map $C$ to the instance $(C,0)$ of $\ApproxOpt(f,m,\varepsilon,\gateset)$. 
    Then $(C,0)$ is a yes-instance iff some circuit $C'$ over $\gateset$ with $f(C')=0$ satisfies $\Metric{C}{C'}{m} \leq \varepsilon$, which is exactly the acceptance condition of $\ApproxMember(\GenericSet_f, m,\varepsilon,\gateset)$.
\end{proof}

\smallskip
\begin{remark}
    \label{rem:factor_approximation}
    Let $f$, $m$, and $\varepsilon$ be as in~\cref{corollary:approx_opt_hardness}, and let $\alpha\geq 1$ be any factor, possibly depending on the input.
    Deciding, given $C$ and $k$, whether there is a circuit $C'$ over $\gateset$ s.t.\ $\Metric{C}{C'}{m}\leq\varepsilon$ and $f(C')\leq\alpha k$ is $\PP$-hard, since the reduction in the proof of~\cref{corollary:approx_opt_hardness} outputs only instances with $k=0$, on which $\alpha k = k$.
\end{remark}

\section{Conclusion and Open Problems} \label{sec:conclusion}

We have shown that approximate quantum circuit optimization is $\PP$-hard for every combination of cost function, metric, and tolerance in~\cref{table:all_problems_hardness2}, and that for total cost functions under any of the Sine, spectral, and operator metrics, and for superposition cost functions under the Sine metric, this covers the entire non-trivial range of $\varepsilon$. 
The results are robust in three directions: they transfer to any distance measure expressible as a strictly increasing function of one of these three, they survive the relaxation from exact minimization to approximation within any factor, and, for trace-based distances and $\GenericSet=\SId$, hardness holds already for classical-reversible gate-sets. 
In particular, within the ranges of~\cref{table:all_problems_hardness2}, relaxing the tolerance does not help, and for the boldface cells no non-trivial tolerance does.

For $\ApproxMember(\SCliff,m,\varepsilon,\gateset)$, our results cover only $0 < \varepsilon < \sqrt{2}/4$, and for $\ApproxMember(\GenericSet,m, \varepsilon, \gateset)$ with $\IGate\in\GenericSet \subseteq \GeneralPermutation$ and $m$ the spectral or operator metric, only $0 < \varepsilon < \sqrt{2}$.
In both cases, the problem becomes trivial only once $\varepsilon$ is large enough that any unitary is $\varepsilon$-close to an element of $\GenericSet$.
We leave open the complexity for $\sqrt{2}/4 \leq \varepsilon$ in the first case and for $\sqrt{2} \leq \varepsilon$ in the second, up to this threshold.
Matching upper bounds are likewise open, with one exception for which our lower bound is tight.
For algebraic $\gateset$ that exactly implements Clifford+$\TGate$ (e.g., Clifford+$\TGate$ itself), $\ApproxIdentity(\SineMetric,\sqrt{2}/2,\gateset)$ is $\PP$-complete, as $\Metric{\IGate}{C}{\SineMetric}^2$ is the acceptance probability of a quantum circuit and $\PQP$ with algebraic amplitudes equals $\PP$~\cite{yamakami2003analysis}.

\section*{Acknowledgements}

\paragraph{Funding.} This work was partially supported by a research grant (VIL42117) from VILLUM FONDEN.

\paragraph{Disclosure of AI use.} An initial draft of the proofs of~\cref{lemma:spectral_GP_distance} and~\cref{theorem:hardness_approx_non_superpos_sp_op} were generated using Claude Opus 5.5 and subsequently checked line by line and rewritten by the authors.
The authors are fully responsible for the whole of this paper.

\bibliographystyle{quantum} 
\bibliography{references}

\appendix 
\section*{Appendix}

\begin{proof}[Proof of~\cref{lemma:hardness_promise_atmostsat}]
    Our proof is by Karp reduction from $\ATMOSTSAT(\beta)$, where we choose $\beta$ s.t. $0<\beta<1$.
    Let a yes/no instance $\varphi$ over $n$ bits be given.
    We define a propositional formula $\varphi'$ s.t. $\FRACSAT{\varphi'} \leq \delta \iff \FRACSAT{\varphi} \leq \beta$, while $\FRACSAT{\varphi'} \leq \kappa$.
    We do so by taking the conjunction of $\varphi$ with a formula $\varphi_1$ over fresh bits that is satisfied by a fraction $c$ of its assignments, where $\delta<c\leq\kappa$, which gives $\FRACSAT{\varphi'} = c\,\FRACSAT{\varphi}$.

    Fix a natural number $L$ with $2^{-L}\leq\kappa-\delta$, which is possible since $\delta<\kappa$, and put
    $$K := \lfloor \kappa 2^{L} \rfloor , \qquad c := K2^{-L} , \qquad \beta := \delta/c \ .$$
    Then $c\leq\kappa\leq 1$ and $c>\kappa-2^{-L}\geq\delta>0$, so $1\leq K\leq 2^{L}$ and $0<\beta<1$.
    Since $\delta$ is $\PTIME$-time computable and $c$ is a fixed dyadic rational, so is $\beta$, and hence $\ATMOSTSAT(\beta)$ is $\PP$-hard (folklore).
    Let $z$ range over $L$ bits, read as an integer in $\{0,\dots,2^{L}-1\}$, and define
    $$\varphi_1(z) := z \leq K-1 \ ,$$
    so that $\FRACSAT{\varphi_1} = K2^{-L} = c$.
    Let $x$ range over the $n$ bits of $\varphi$, and define
    $$\varphi'(x,z) := \varphi(x) \wedge \varphi_1(z)$$
    as the output of the reduction.

    We argue that the reduction takes polynomial time.
    The constants $L$ and $K$ depend only on $\delta$ and $\kappa$, so $\varphi_1$ is a fixed formula over $L$ bits and can be written over $\vee$, $\wedge$, and $\neg$ with constant size (e.g., in disjunctive normal form).
    Hence $\varphi'$ is computed from $\varphi$ in linear time.

    For correctness, note that $\varphi'$ ranges over $n+L$ bits and is satisfied by exactly the pairs $(x,z)$ with $\varphi(x)$ and $\varphi_1(z)$, so
    $$\FRACSAT{\varphi'} = \frac{2^{n}\FRACSAT{\varphi}\cdot 2^{L}\FRACSAT{\varphi_1}}{2^{n+L}} = c\,\FRACSAT{\varphi} \ .$$
    Hence, as $c>0$,
    $$\FRACSAT{\varphi'}\leq\delta \iff \FRACSAT{\varphi}\leq\delta/c = \beta \ ,$$
    so $\varphi'$ is a yes-instance of $\PROMISEATMOSTSAT(\delta,\kappa)$ iff $\varphi$ is a yes-instance of $\ATMOSTSAT(\beta)$.
    The promise holds since $\FRACSAT{\varphi'} = c\,\FRACSAT{\varphi} \leq c \leq \kappa$, where we use $\FRACSAT{\varphi}\leq 1$.
\end{proof}

\end{document}